\documentclass[sigconf]{acmart}
\PassOptionsToPackage{prologue,dvipsnames}{xcolor}
\usepackage{graphicx}
\usepackage{graphicx}
\usepackage{multirow}
\usepackage{booktabs}
\usepackage{amsmath}
\usepackage{subfig}

\setcopyright{none}
\AtBeginDocument{%
  }

\copyrightyear{2025}
\acmYear{2025}
\setcopyright{acmlicensed}\acmConference[SIGIR '25]{Proceedings of the 48th International ACM SIGIR Conference on Research and Development in Information Retrieval}{July 13--18, 2025}{Padua, Italy}
\acmBooktitle{Proceedings of the 48th International ACM SIGIR Conference on Research and Development in Information Retrieval (SIGIR '25), July 13--18, 2025, Padua, Italy}
\acmDOI{10.1145/3726302.3729976}
\acmISBN{979-8-4007-1592-1/2025/07}

\usepackage{verbatim}
\usepackage{graphicx} 
\usepackage{float} 
\usepackage{multirow}
\usepackage{fancyhdr}
\usepackage{enumitem}
\usepackage{xspace}
\usepackage{colortbl}

\newcommand{\ie}{\emph{i.e., }}

\begin{document}

%\title{Smart Fitting Room: A Generative Approach to Matching-aware Virtual Try-On}

\title{FashionDPO: Fine-tune Fashion Outfit Generation Model using Direct Preference Optimization 
}


\author{Mingzhe Yu}
\email{mz_y@mail.sdu.edu.cn}
\affiliation{%
  \institution{Shandong University}
  \city{Jinan}
  \country{China} 
}

\author{Yunshan Ma}
\email{ysma@smu.edu.sg}
\affiliation{%
  \institution{Singapore Management University}
  \country{Singapore}
}

\author{Lei Wu}
\email{i_lily@sdu.edu.cn}
\authornote{Corresponding author. This research was supported by the Singapore Ministry of Education (MOE) Academic Research Fund (AcRF) Tier 1 grant (No. MSS24C018) and the Excellent Young Scientists Fund Program of Shandong Province (Grant No.ZR2024QF251).}
\affiliation{%
  \institution{Shandong University}
  \city{Jinan}
  \country{China}
}

\author{Changshuo Wang}
\email{cswang02@outlook.com}
\affiliation{%
  \institution{Shanghai Jiao Tong University}
  \city{Shanghai}
  \country{China}
}

\author{Xue Li}
\email{xue.lii754@gmail.com}
\affiliation{%
  \institution{Jiangnan University}
  \city{Wuxi}
  \country{China}
}

\author{Lei Meng}
\email{lmeng@sdu.edu.cn}
\affiliation{%
  \institution{Shandong University}
  \city{Jinan}
  \country{China}
}

\renewcommand{\shortauthors}{Mingzhe Yu et al.}


\begin{abstract}

% Fashion plays a crucial role in showcasing individual expression and character, but many individuals may lack the expertise to navigate rapidly evolving trends, making outfit coordination challenging. To address this, Outfit Recommendation (OR) systems have been widely applied. Recent advancements in generative models enable the direct recommendation of high-quality fashion items. However, existing generative models rely on supervised learning, which mismatches the task of fashion recommendation due to the inherent variability in suitable fashion items for incomplete outfits. To overcome this limitation, we propose FashionDPO, a generative fashion recommendation model based on direct preference feedback optimization. Our model shifts the learning objective from ground truth to preference feedback, enabling it to generate diverse and innovative fashion products. We introduce a comprehensive expert feedback system, evaluating generated outfits on quality, compatibility, and personalization, which fine-tunes the model using the Direct Preference Optimization (DPO) method. Experiments on iFashion and Polyvore-U datasets demonstrate the effectiveness of our approach in generating fashion products that align with user preferences while adhering to design principles.

Personalized outfit generation aims to construct a set of compatible and personalized fashion items as an outfit. Recently, generative AI models have received widespread attention, as they can generate fashion items for users to complete an incomplete outfit or create a complete outfit. However, they have limitations in terms of lacking diversity and relying on the supervised learning paradigm. 
Recognizing this gap, we propose a novel framework FashionDPO, which fine-tunes the fashion outfit generation model using direct preference optimization.
This framework aims to provide a general fine-tuning approach to fashion generative models, refining a pre-trained fashion outfit generation model using automatically generated feedback, without the need to design a task-specific reward function.
To make sure that the feedback is comprehensive and objective, we design a multi-expert feedback generation module which covers three evaluation perspectives, \ie quality, compatibility and personalization.
Experiments on two established datasets, \ie iFashion and Polyvore-U, demonstrate the effectiveness of our framework in enhancing the model's ability to align with users' personalized preferences while adhering to fashion compatibility principles.
Our code and model checkpoints are available at \url{https://github.com/Yzcreator/FashionDPO}.

\end{abstract}

\ccsdesc[500]{Information systems~Multimedia and multimodal retrieval}
\ccsdesc[500]{Information systems~Recommender systems}
\ccsdesc[500]{Information systems~Multimedia content creation}
% \ccsdesc[100]{Computing methodologies~Machine learning}

%\keywords{Generative Recommendation, Personalization, Direct Preference Optimization}

\keywords{Fashion Outfit Generation, Fashion Image Generation,  Generative Fashion Recommendation}
%, Generative Recommendation, Multimodal Recommendation}


%   \includegraphics[width=1.0\textwidth]{image/front page.pdf}
%   \setlength{\abovecaptionskip}{0.1cm} 
%   \setlength{\belowcaptionskip}{0.cm} 
%   \caption{Mix-and-match and try-on are two essential fashion needs in daily life. Instead of separately modeling, we propose a one-stop system of Smart Fitting Room, in which the user only needs to provide a partially masked image as query, our system will generate an apparel to match with the query and put it on the query image.}
%   \label{fig:frontpage}


\maketitle

\begin{figure}
  \includegraphics[width=0.47\textwidth]{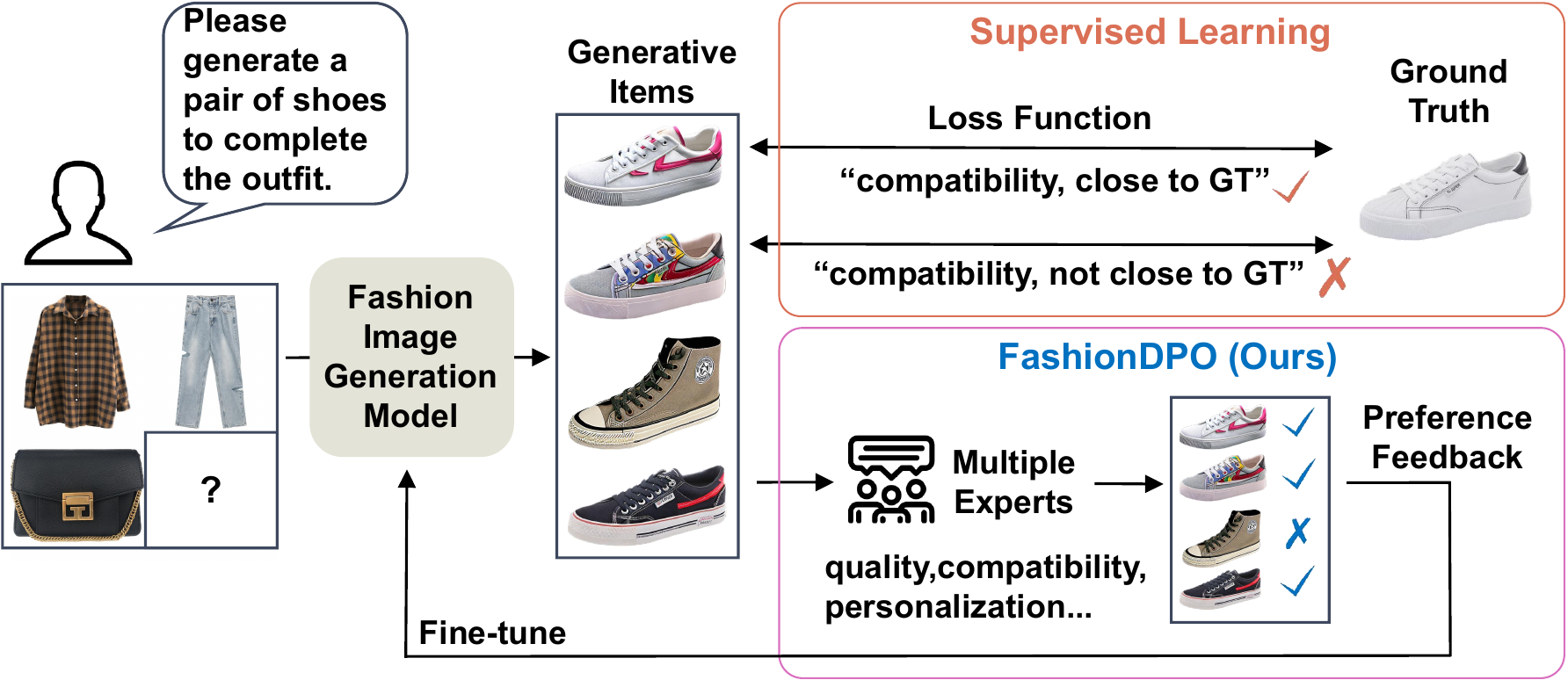}
  \setlength{\abovecaptionskip}{-0.2cm} 
  \setlength{\belowcaptionskip}{-0.4cm} 
  \caption{Illustration of our motivation and the paradigm comparison between FashionDPO and the supervised learning methods, where FashionDPO optimizes the model using feedback from multiple experts without relying on labeled dataset, resulting in high diversity while maintaining high quality.}
  \label{Intro-DPO}
\end{figure}

\section{Introduction}

Fashion in life is not merely about clothing and dressing, while it serves as a means of expressing personal style, identity, and lifestyle ~\cite{FashionRecSurvey-23}. 
One of the major requirements in people's daily fashion life is personalized fashion matching, where people need to pick up a set of fashion items (defined as an outfit) that are compatible with each other and fit his/her personalized fashion taste ~\cite{DiFashion, HMaVTON}. In addition to experts in the domain of fashion, researchers in information retrieval also express interest in studying this problem, which is defined as the task of personalized outfit recommendation ~\cite{POG, PORAnchors, PFOG}. 
%The specific tasks include obtaining a compatible fashion item based on the user's interaction history to fill an incomplete outfit—referred to as Personalized Fill-in-the-Blank (PFITB), and assembling a compatible, complete outfit for personalized recommendation.
The specific tasks include 1) Personalized Fill-in-the-Blank (PFITB), completing an incomplete outfit by finding a compatible fashion item based on the user's interaction history, and 2) Generative Outfit Recommendation (GOR), assembling a complete outfit from scratch catering to both personalized preference and fashion item compatibility.

Regarding these specific tasks, existing works can be categorized into retrieval-based methods and generation-based methods.
The retrieval-based models~\cite{POG,PFOG,A-FKG,withLSTM} typically explore the existing candidate item set and select certain items to either complete a partial outfit or curate an outfit from scratch. 
Despite their prevailing usage, such retrieval-based methods are inevitably restricted by the limited inclusivity and diversity of existing items. ~\cite{Compatibility, HMaVTON} What if we cannot find a suitable item after iterating every item within the available item set?
To address this problem, a natural and reasonable solution is to directly generate an image of the desired fashion item using the cutting-edge image generation models~\cite{CRAFT, DiFashion, OutfitGAN,SD, controlNet}. For example, DiFashion~\cite{DiFashion} leverages the pre-trained stable diffusion model and incorporates users' historical preference and items' compatible constraints, achieving SOTA performance on both tasks of PFITB and GOR.  
Despite its promising performance, we pinpoint a problem that the generated fashion items exhibit limited diversity, which is probably due to the fact that the pre-defined outfits in the dataset cannot fully cover all possible fashion item combinations.
% thereby hindering the model’s ability to learn the various styles and preferences.
For example, in the PFITB task, each incomplete outfit is paired with only one ground-truth item, which, however, may have multiple alternative items that can also satisfy the personalization and compatibility constraints.
Furthermore, current generation-based methods solely rely on the supervised learning paradigm, which exacerbates the issue of lacking diversity during the generation ~\cite{Diffusion-DPO, D3PO}. 
As shown in Figure \ref{Intro-DPO}, during the training process, the generative models ~\cite{OutfitGAN,DiFashion} aim to minimize the difference between the result image and the ground-truth image. Thereby, when the generated images visually adhere to fashion compatibility but differ from the ground truth, the model would erroneously penalize these deviations. Consequently, the model would overly focus on specific features within the pre-defined outfits, rather than learning more generalized principles of clothing compatibility, resulting in limited generation diversity.

To resolve the limitations of lacking diversity and supervised learning paradigm, intuitive solutions could be augmenting the dataset through additional human annotations. 
However, relying on human annotations yield high monetary cost and time expense, even worse, the involvement of human annotators would introduce uncontrollable subjective biases into the dataset. Additionally, it becomes challenging to adapt or update the dataset when the fashion trends are constantly changing and evolving ~\cite{fashion_trend,fashion_trend_tmm}. 
Considering the above concerns, a more advanced and promising approach is the RLAIF~\cite{RLAIF-sample} (reinforcement learning from AI feedback), which enables the model to learn from a variety of possible feedback and adapt to more diverse situations. This method is particularly suitable for scenarios, such as artistic creation, where no fixed answers exist. Nevertheless, how to design the AI models to ensure that the feedback is both comprehensive and objective still remains a challenge. 
Moreover, how to leverage the feedback generated by AI models to guide the training of generation models is still challenging and under-explored in the fashion domain. 
For example, existing methods~\cite{RLAIF} typically first train a reward model that is aligned with human preferences and then fine-tune the generation model using reinforcement learning. However, constructing an effective reward model demands large-scale datasets and multiple rounds of parameter tuning, which is time-consuming and resource-intensive.

To tackle these challenges, we propose a novel framework FashionDPO, which fine-tunes the fashion outfit generation model using direct preference optimization. 
To make sure the automatic feedbacks are comprehensive and professional, we design a multi-expert feedback generation module to address the first challenge, which consists of three experts covering diverse evaluation perspectives: 1) Quality: evaluating whether the fashion elements in the image are complete and conform to fashion design principles, 2) Compatibility: assessing whether the generated fashion products are coordinated within an outfit, and 3) Personalization: ensuring that the recommended fashion products align with the user's personal preferences. To solve the second challenge, we leverage the DPO (Direct Preference Optimization)~\cite{DPO} framework and group the feedbacks into positive-negative pairs, which are collected from multiple experts' feedback. We then utilize the positive-negative pairs to guide the fine-tuning process, eliminating the need to train a reward model. In summary, \textbf{the primary contributions} of our work are as follows:

\begin{itemize}[leftmargin=0.5cm, itemindent=0cm]
\item We identify the limitations of lacking diversity and supervised learning paradigm in existing personalized outfit generation models, and we propose to utilize mulitple AI feedbacks to solve these limitations.

% \item We identify the limitations of lacking diversity and supervised learning paradigm in existing personalized outfit generation models and propose to leverage AI feedbacks to address the problem.

\item Fulfilling the idea of using AI feedbacks, we propose a novel framework FashionDPO, which fine-tunes the generation model using DPO based on multiple AI experts' feedbacks.

\item Extensive experiments on iFashion and Polyvore-U datasets demonstrate the effectiveness of our framework regarding various evaluating metrics.
\end{itemize}

\section{Related Work}

% 先介绍个性化的时尚穿搭推荐作品，然后再进入生成作品。
% From Introduction: However, some people may lack professional fashion knowledge, leading to confusion when confronted with rapidly changing fashion trends, making it difficult for them to put together compatible outfits that suit their style ~\cite{OGR}. Consequently, Outfit Recommendation (OR) has gained widespread application in the fashion domain ~\cite{PORGraph, PORAnchors, A-FKG}. Meanwhile, the rapid development of generative models ~\cite{DDIM, SD, controlNet, lora} has made it possible to directly recommend high-quality generated fashion products to users. 

\noindent \textbf{Fashion Outfit Recommendation.}
In the fashion domain, Outfit Recommendation (OR)~\cite{PORGraph, PORAnchors, A-FKG} has gained widespread application. There are two requirements in fashion outfit recommendation: compatibility and personalization. Furthermore, it is also a popular task in the domain of computational fashion ~\cite{FashionRecSurvey-23}. 
Early works ~\cite{personalCom, PORAnchors, A-FKG} primarily focused on compatibility, aiming to retrieve already well-matched outfits for users.
Some works ~\cite{PFOG, POG} attempt to introduce personalization in the recommendation process, combining a set of personalized and compatible items into an outfit that aligns with fashion styling principles. 
Moreover, bundle recommendation, a more generalized recommendation paradigm, subsume personalized fashion outfit recommendation as one of its applications. Multiple works~\cite{MultiCBR,EBRec,BundleMLLM} have been proposed by using graph learning, contrastive learning, as well as multimodal large language models. 
Despite various progress, the above works follow the retrieval paradigm and are constrained by the variety and quantity of fashion products in the dataset, making it difficult to meet users' personalized needs, especially in terms of texture and other details. However, with the rapid development of generative models ~\cite{SD, controlNet, lora}, the quality and diversity of image generation have significantly improved, making it possible to directly recommend generated custom fashion products to users. Recent work ~\cite{DiFashion} has introduced the PFITB task, which combines the user's interaction history with fashion products to generate a personalized matching outfit.

%However, the limited quantity and variety of clothing in the dataset prevent it from meeting users' personalized needs ~\cite{PFOG}. 

\noindent \textbf{Fashion Image Generation.} It refers to the task of generating fashion-related images using deep learning models. This task is widely applied in the fashion domain, covering areas such as clothing design, virtual try-on, and personalized recommendation, among others~\cite{yang2018recommendation, Compatibility,FashionReGen24}. Previous works, such as CRAFT~\cite{CRAFT}, generate feature representations for clothes pairings and retrieve the most suitable individual clothes items from the dataset. In the virtual try-on domain, previous works ~\cite{VITON, GP-VTON} based on GANs involve generating warped clothes aligned with character, and then generating images of the character wearing the warped clothes. The diffusion models ~\cite{DCI-VTON} enhance image quality by replacing the generator in the second stage. Current work ~\cite{stableVTON} learns the semantic correspondence between the clothing and the human body within the latent space of the pre-trained diffusion model in an end-to-end manner.
In the personalized recommendation domain, HMaVTON ~\cite{HMaVTON} generates diverse and well-matched fashion items to the given person. Existing personalized image generation models ~\cite{Jedi, ELITE, PathchDPO, BDPO} aim to generate images aligned with reference styles or elements, yet recommending images consistent with a user's interaction history is meaningless.
% And DVBPR ~\cite{DVBPR} generates clothes images based on user preferences but is limited to generating images that are identical in shape to those in the dataset. 

\begin{figure*}[ht]
  \centering
  \setlength{\abovecaptionskip}{0.1cm} 
  \setlength{\belowcaptionskip}{-0.3cm}
  \includegraphics[width=1.0\textwidth]{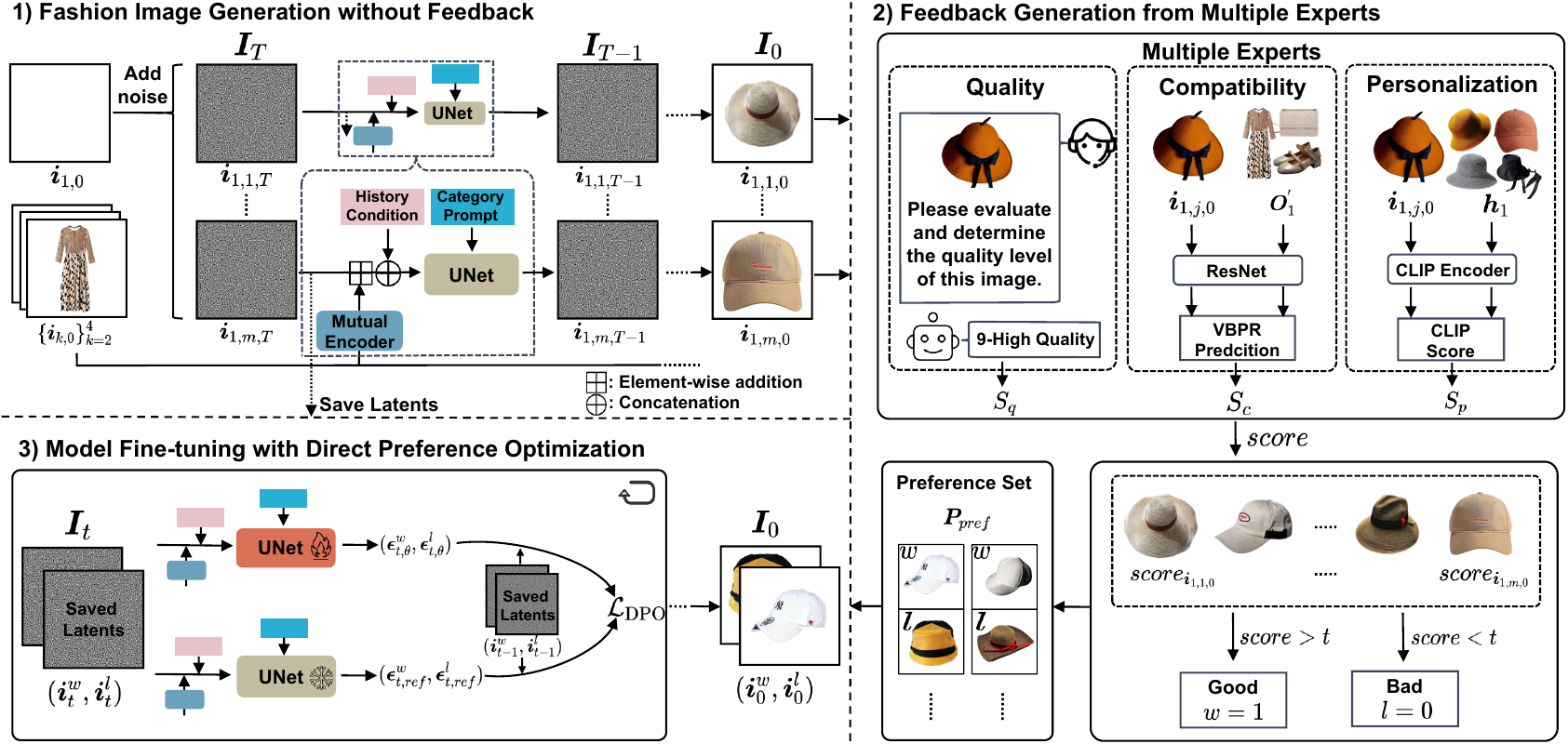} 
  \caption{
   The overview of FashionDPO, which consists of three consecutive key modules: 1) Fashion Image Generation without Feedback, 2) Feedback Generation from Multiple Experts, and 3) Model Fine-tuning with Direct Preference Optimization.  
   %We design comprehensive AI models to evaluate generated fashion items and update the model parameters using the preference information contained in the AI feedback. 
  }
  \label{fig:model}
\end{figure*}

% 补充：
% sd,controlnet应用于clothing design
% try-on, 两阶段gan+dm，一阶段dm

\noindent \textbf{Direct Preference Optimization.}
In the field of natural language processing, Direct Preference Optimization (DPO) has been proposed to reduce
training costs ~\cite{DPO}, which uses preferences rather than explicit rewards to fine-tune LLMs. This approach is also applied to the post-training of text-to-image diffusion models. Diffusion-DPO ~\cite{Diffusion-DPO} fine-tunes the generative model in a single step after receiving feedback from the Preference Evaluator. D3PO ~\cite{D3PO} assumes that the preferred outcome holds true for all time steps in the diffusion model and fine-tunes each of the time steps in the generative model based on the feedback results. It demonstrates that in diffusion models, directly updating the policy based on human preferences within an MDP is equivalent to first learning the optimal reward model and then using it to guide policy updates. SPO ~\cite{SPO} assesses preferences at each time step during the sampling process and adds noise to the preferred image to generate the noise image for the next time step. We introduce DPO into generative fashion recommendation, where learning based on preference feedback eliminates the constraints of ground truth, showcasing richer possibilities in clothing textures and details.

% Problem:imited diversity, misalignment between tasks and objectives
% Challenge:%The challenge is the design of AI evaluation models and the mechanisms for updating feedback.

\section{Preliminary}
We first introduce the problem formulation, 
followed by a briefing of the diffusion model and the direct preference optimization.

% In the fine-tune process, we use the preference information contained in the binary pairs to guide the fine-tuning process, iteratively incorporating fashion knowledge into the model.
% Through this fine-tuning framework, the model is freed from the constraints of ground truth on the output products, enabling it to explore a broader range of matching capabilities.
% we interpret the diffusion process in generation models as a multi-step Markov Decision Process (MDP). Based on the feedback from the multiple experts, we extend Diffusion Process Optimization (DPO) into a multi-step MDP.

\noindent \textbf{Problem Formulation of PFITB and GOR.} Based on the user information $\boldsymbol{u}$, these tasks aim to generate complete outfit $\boldsymbol{O} = \{\boldsymbol{i}_k\}_{k=1}^n$, where $\boldsymbol{O}$ denotes complete outfit and $\boldsymbol{i}_k$ is individual fashion item. 
Specifically, the PFITB task generates compatible fashion items $\boldsymbol{i}_k$ for each incomplete outfit $\boldsymbol{O}^{'} = \boldsymbol{O} \backslash \{\boldsymbol{i}_k\}$ according to user preferences. And the GOR task is further expanded to generate a complete set of matching outfits $\boldsymbol{O}$ for users.

\noindent \textbf{Diffusion Models.} Diffusion Models learn probability distribution $p(\boldsymbol{i})$ by inverting a Markovian forward process $q(\boldsymbol{i}_t|\boldsymbol{i}_{t-1})$. The GOR task, starting from multiple randomly initialized noisy images $\boldsymbol{O}_T = \{\boldsymbol{i}_{k,T}\}_{k=1}^n$ where $\boldsymbol{i}_{k,T} \sim \mathcal{N}(0,I)$, the diffusion models gradually remove noise from the image through a reverse process. At time step $t \in \{1,...,T\}$, the U-Net predicts the noise $\boldsymbol{\epsilon}_k$ of the $\boldsymbol{i}_{k,t}$, ultimately generating personalized fashion outfits $\boldsymbol{O}_0 = \{\boldsymbol{i}_{k,0}\}_{k=1}^n$.
And the PFITB task starts with a single noisy image $\boldsymbol{i}_{k,T}$ and generates a fashion item $\boldsymbol{i}_{k,0}$ for incomplete outfit $\boldsymbol{O}^{'}$.
To achieve personalized outfit generation, we introduce the user's interaction history $\boldsymbol{h}_k$, category prompt $\boldsymbol{p}_k$ and mutual influence $\boldsymbol{m}_k$ between items in the outfit as condition $\boldsymbol{c}_k$. In the case of conditional generative modeling, the U-Net learns the probability distribution $p(\boldsymbol{i}|\boldsymbol{c}_k)$ with the following objective:
\begin{equation}
    \mathcal{L} = \mathbb{E}_{t,\boldsymbol{\epsilon}_k \sim \mathcal{N}(0,I)} [||\boldsymbol{\epsilon}_k - \boldsymbol{\epsilon}_{\theta}(\boldsymbol{i}_{k,t},\boldsymbol{h}_k,\boldsymbol{p}_k,\boldsymbol{m}_k,t)||^2_2].
\end{equation}

\noindent \textbf{Direct Preference Optimization.} The purpose of preference-Based Models is to optimize the model's output according to the user's preferences. It often requires learning a reward function $r:\mathcal{S} \times \mathcal{A} \rightarrow \mathbb{R}$  from human feedback, where $\mathcal{S},\mathcal{A}$ denotes state space and action space. The objective is to find a policy $\pi(a|s)$ that maximizes the reward $r$, where $a \sim \mathcal{A}, s \sim \mathcal{S}$.
However, user preferences are typically complex and multi-dimensional, making it a challenge to accurately model and design the corresponding reward function. Recently, the direct preference optimization (DPO) ~\cite{DPO} has been proposed to fine-tune models using direct feedback from users. Given a state $s$ and action pairs $(a_w,a_l)$, the $\pi_{\theta}(a|s)$ represents the sampling policy of the finetuned model, and $\pi_{ref}(a|s)$ represents the policy of the reference model. In the GOR task, we can generate fashion pairs $(y_w,y_l)$ based on prompt $x$, where $y_w$ is user's preference choice and $y_l$ is the inferior one. LPO formulates a maximum likelihood objective for a parametrized policy $\pi_\theta$:
{\small
\begin{equation} \label{dpoloss}
\mathcal{L} =-\mathbb{E}_{(x,y_{w},y_{l})\sim\mathcal{D}}\left[\log\sigma\left(\beta_1\log\frac{\pi_{\theta}(y_{w}|x)}{\pi_{\mathrm{ref}}(y_{w}|x)}-\beta_2\log\frac{\pi_{\theta}(y_{l}|x)}{\pi_{\mathrm{ref}}(y_{l}|x)}\right)\right],
\end{equation}
}
where $\beta$ is used to control the deviation between $\pi_{\theta}$ and $\pi_{ref}$. 

% {\scriptsize
% \begin{equation} \label{dpoloss}
% \mathcal{L}_{\mathrm{DPO}}(\pi_{\theta};\pi_{\mathrm{ref}})=-\mathbb{E}_{(x,y_{w},y_{l})\sim\mathcal{D}}\left[\log\sigma\left(\beta_1\log\frac{\pi_{\theta}(y_{w}|x)}{\pi_{\mathrm{ref}}(y_{w}|x)}-\beta_2\log\frac{\pi_{\theta}(y_{l}|x)}{\pi_{\mathrm{ref}}(y_{l}|x)}\right)\right],
% \end{equation}
% }

\section{Method}
We present our proposed method FashionDPO, which consists of three consecutive modules: 1) fashion image generation without feedback, 2) feedback generation from multiple experts, and 3) model fine-tuning with direct preference optimization. We explicate the detailed design of these modules one by one.

%We first generate multiple fashion item images for the incomplete outfit as a candidate set for evaluation.
% , which gets rid of the constraints of the ground truth.
%In the feedback process, we design a multi-expert feedback generation module to evaluate items, which provides feedback as binary pairs. 
%In the fine-tune process, we use the binary pairs to guide the fine-tuning process, incorporating fashion knowledge into the model.
% It does not need a reward model; instead, it optimizes the model parameters based on the preference results from feedback.
% To obtain preference feedback, we first generate fashion items by considering user information, \ie user's interaction history. This process is repeated multiple times to generate several sets of data for comparison.
% Then we design an evaluation framework to gather feedback from multiple evaluators on various aspects of the model, \ie image quality, compatibility, and personalization. And we apply the DPO theory to fashion outfit generation models, incorporating feedback and preference information into the model.

\subsection{\textbf{Fashion Image Generation without Feedback}} \label{inference}
We faithfully follow DiFashion~\cite{DiFashion} to build the generation model.
Starting from empty image $\boldsymbol{i}_{1,0}$ incomplete outfit $\boldsymbol{O}^{'} = \{\boldsymbol{i}_{k,0}\}_{k=2}^n$, where the $\boldsymbol{i}_{k,0}$ represents a fashion item in the outfit. During the initialization process, the model adds random Gaussian noise to image $\boldsymbol{i}_{1,0}$. The forward formula is expressed as follows:
\begin{equation}
    \boldsymbol{i}_{k,t} = \sqrt{\alpha_t} \boldsymbol{i}_{k,0} + \sqrt{1-\alpha_t}\boldsymbol{\epsilon}_k,
\end{equation}
where $\boldsymbol{\epsilon}_k \sim \mathcal{N}(0, 1)$ is Gaussian noise sampled from a standard normal distribution, and $\alpha_t$ is a decay coefficient with time step $t \in \{1,2,...,T\}$. As $t$ increases, the influence of noise $\boldsymbol{\epsilon}$ gradually grows, and $\boldsymbol{i}_{k,T}$ becomes almost the entire noise. To obtain multiple generated results for extracting preference information, we apply the forward noising process $m$ times to obtain $\boldsymbol{I}_T = \{\boldsymbol{i}_{k,j,T}\}_{j=1}^m$.

In the reverse process, the model starts with noisy images $\boldsymbol{i}_{k,j,t}$ as input, gradually predicts and removes the noise step by step, and ultimately reconstructs the fashion items $\boldsymbol{i}_{k,j,0}$. The denoising process is outlined as follows:
               
\begin{equation}
    \boldsymbol{i}_{k,j,0} = \frac{1}{\sqrt{\alpha_t}} (\boldsymbol{i}_{k,j,t} - \sqrt{1-\alpha_t}\boldsymbol{\epsilon}_\theta(\boldsymbol{i}_{k,j,t},t,\boldsymbol{c}_k)),
\end{equation}
where the $\boldsymbol{\epsilon}_\theta(\boldsymbol{i}_{k,j,t},t,\boldsymbol{c}_k)$ represents the noise predicted by the model, and $\boldsymbol{c}_k$ denotes the condition. At each time step $t$, the model first saves the current latent images for the subsequent fine-tuning stage. As shown in Figure \ref{fig:model}, the model introduces the mutual influence $\boldsymbol{m}_k$, user's interaction history $\boldsymbol{h}_k$ and category prompt $\boldsymbol{p}_k$ as condition $\boldsymbol{c}_k$: 

\textbf{Mutual Influence.}
To ensure that the generated fashion items within an outfit are well-matched, the model introduces mutual influence $\boldsymbol{m}_k$ during the inference process, enabling the fashion items to influence each other. 
At time step $t$, the model uses the incomplete outfit $\boldsymbol{O}^{'} = \{\boldsymbol{i}_{k,0}\}_{k=2}^4$ to obtain the latent feature representation $\boldsymbol{m}_{1,t}$ through the Multi-Layer Perceptron (MLP), providing outfit matching information for $\boldsymbol{i}_{1,j,t}$. And it fuses two latent representations $\boldsymbol{m}_{1,t}$ and $\boldsymbol{i}_{1,j,t}$ through element-wise addition:
\begin{equation}
    \boldsymbol{i}_{1,j,t}^m = (1-\eta) \cdot \boldsymbol{i}_{1,j,t} + \eta \cdot \boldsymbol{m}_{1,t},
\end{equation}
where $\eta$ is a hyperparameter that controls the influence of mutual condition. A larger $\eta$ indicates a greater impact on the noisy image.

\textbf{User’s Interaction History.} 
The personalized information of the user is reflected through their fashion items' interaction history. For user $\boldsymbol{u}$, the fashion items in the interaction history that belong to the same category as $\boldsymbol{i}_{1,0}$ are extracted to form $\boldsymbol{u}_k = \{\boldsymbol{i}_{k}^h\}_{h=1}^N$. And the model use the pre-trained autoencoder $\mathcal{E}$ in diffusion models~\cite{SD} to encode the user’s interaction history images $\boldsymbol{u}_k$ and take the average as the history condition $\boldsymbol{h}_k$:
\begin{equation}
    \boldsymbol{h}_k = \frac{1}{N} \sum_{h=1}^N \mathcal{E}(\boldsymbol{i}_{k}^h).
\end{equation}
Then the model concatenates $\boldsymbol{i}_{1,j,t}^m$ and $\boldsymbol{h}_k$ as input to the U-Net.

\textbf{Category Prompt.} In order to make the model's generated results rely more on its learned fashion matching knowledge rather than the provided prompt, the model uses a brief text description that contains only the relevant fashion categories. For example, in Figure \ref{fig:model}, it use "A photo of a hat, with the white background" as prompt and use the CLIP text encoder~\cite{CLIP} to encode text and obtain the Category Prompt $\boldsymbol{p}_k$.

In summary, at each sampling time step $t$, the model fuses the features of the mutual condition into the noisy image and concatenates the image features from the user interaction history. These combined latent features are then fed into UNet for noise prediction.

% In an outfit, we will generate fashion item $\boldsymbol{i}_k$ based on the incomplete outfit $\boldsymbol{O}_k^{'} = \boldsymbol{O} \backslash \{\boldsymbol{i}_k\}$. To facilitate preference-based comparisons later, we execute this process in parallel $j$ times to obtain $\boldsymbol{I}_0 = \{\boldsymbol{i}_{k,j}\}_{j=1}^m$. Specifically,  the forward stochastic noising process to obtain I

\subsection{\textbf{Feedback Generation from Multiple Experts}}
From the above inference process, we obtain multiple generated fashion items $\boldsymbol{I}_0 = \{\boldsymbol{i}_{k,j,0}\}_{j=1}^m$. To ensure a comprehensive and objective evaluation, we designed a framework involving multiple experts to evaluate $\boldsymbol{I}_0$ from three aspects: quality, compatibility, and personalization. The feedback is then utilized to fine-tune the model at each iteration. Next, we introduce three evaluation perspectives and their corresponding expert models:
% Notably, for each evaluation aspect mentioned above, we have selected a specific model as the evaluation criterion. 

% Different expert models may have varying impacts on feedback and fine-tuning processes, and we will further analyze this in the Section \ref{expert implementation}. 

\textbf{Quality.}
When evaluating the quality of generated images, traditional metrics such as FID~\cite{FID} and LPIPS~\cite{LPIPS} typically capture only low-level statistical features and fail to fully reflect the complex semantic information perceived by humans. Therefore, we use the multimodal large model MiniCPM~\cite{minicpm} to score the images. We input the generated fashion items $\boldsymbol{I}_0$ into the model and define the task as an image quality classification problem. Specifically, we categorize image quality into ten levels, ranging from "1-Poor Quality" to "10-Exceptional Quality," and provide the following evaluation criteria: "Consider whether the fashion elements in the image are complete and whether they conform to fashion design principles." The MiniCPM is responsible for determining which quality level the input image $\boldsymbol{i}_{k,j,0}$ belongs to and providing a response accordingly. By analyzing the image quality classification levels in MiniCPM's response, we can derive the image quality score $S_q = \{s_q^j\}_{j=1}^m$.

\textbf{Compatibility.}
We follow one of the typical methods VBPR~\cite{VBPR} to build the compatibility score prediction module. For example, given the generated fashion item $\boldsymbol{i}_{1,1,0}$ and its corresponding incomplete outfit $\boldsymbol{O}_1^{'}$, we employ the pre-trained deep CNN model, \ie ResNet-50~\cite{resnet} to extract the visual features and then leverage a linear layer to transform the visual features into a shared representation space, which is formally defined as:
\begin{equation}
\begin{aligned}
  \boldsymbol{v}_i &= \text{ResNet}(\boldsymbol{i}_{1,1,0})\mathbf{W}_1, \\
  \boldsymbol{v}_o &= \text{ResNet}(\boldsymbol{O}_{1}^{'}) \mathbf{W}_1 =\frac{1}{3} \sum_{k=2}^4 \text{ResNet}(\boldsymbol{i}_{k,0})\mathbf{W}_1,
  % \vspace{-0.2cm}
\end{aligned}
\end{equation}
where $\boldsymbol{v}_i, \boldsymbol{v}_o \in \mathbb{R}^d$ are visual representations, $\mathbf{W}_1 \in \mathbb{R}^{2048 \times d}$ is the feature transformation matrix of the linear layer, $\text{ResNet}(\cdot)$ represents the ResNet-50 network, and $d$ is the dimensionality of the visual representation. Thereafter, we can calculate the compatibility score $\boldsymbol{s}_{i,o}$ via the following equation:
\begin{equation}~\label{eq:compatibility_score}
  s_{i,o} = \alpha + \beta_i + \beta_o + \boldsymbol{v}_i^{\intercal}\boldsymbol{v}_o,
\end{equation}
where $\alpha, \beta_i, \beta_o$ are trainable parameters to model the bias for the global, item $\boldsymbol{i}$, and incomplete outfit $\boldsymbol{O}^{'}$. We use Bayesian Personalized Ranking (BPR)~\cite{bpr} loss to train the model, denoted as:
\begin{equation}
  \mathcal{L}^{\text{BPR}} = \sum_{(i,r,o) \in \mathcal{Q}}{-\text{log}\sigma(s_{i,o}-s_{r,o})},
\end{equation}
where $\mathcal{Q}=\{i,r,o)|\boldsymbol{s}_{i,o}=1, \boldsymbol{s}_{r,0}=0\}$, $\boldsymbol{s}_{i,o}$ is the ground-truth matching relation between item $\boldsymbol{i}$ and incomplete outfit $\boldsymbol{O}^{'}$, $\boldsymbol{s}_{i,o}$ indicates that $(\boldsymbol{i},\boldsymbol{O}^{'})$ are matched with each other. In contrast, $s_{r,o}=0$ means the random item $\boldsymbol{r}$ and $\boldsymbol{O}^{'}$ are an unmatched pair. And $\sigma(\cdot)$ represents the sigmoid function.
The generated fashion items $\boldsymbol{I}_0 = \{\boldsymbol{i}_{k,j,0}\}_{j=1}^m$ received a set of scores $S_c = \{s_{i,o}^j\}_{j=1}^m$ after being evaluated by compatibility assessment expert.

\textbf{Personalization.}
The user's interaction history with fashion items reflects their preferences. We designed a personalization evaluation expert to assess whether the generated results  satisfies these preferences. Specifically, We use a pretrained CLIP image encoder to encode the generated fashion item $\boldsymbol{i}_{k,j,0}$, obtaining $\boldsymbol{v}_k$, and mapping it to the same latent space as the history condition $\boldsymbol{h}_k$. Then, we measure the similarity between them by calculating their cosine similarity:
\begin{equation}
    s_{p} = \text{CLIP\_Score}(\boldsymbol{v}_k, \boldsymbol{h}_k) = \frac{\boldsymbol{v}_k \cdot \boldsymbol{h}_k}{\parallel\boldsymbol{v}_k\parallel \parallel\boldsymbol{h}_k\parallel}.
\end{equation}
Similarly, we can obtain the personalization score $S_p = \{s_{p}^j\}_{j=1}^m$. Then we combine the evaluation results of multiple experts to obtain a weighted score:
\begin{equation}
    score = \alpha_q \cdot \text{norm}(S_q) + \alpha_c \cdot \text{norm}(S_c) + \alpha_p \cdot \text{norm}(S_p),
\end{equation}
where the $\alpha_q, \alpha_c, \alpha_p$ is a hyperparameter that controls the weight of expert evaluation, and $\text{norm}(\cdot)$ denotes the min-max normalization. We compare the score with the threshold $\boldsymbol{t}$: if $score_{\boldsymbol{i}_{k,j,0}} > t$, it indicates that the generated item meets the design and personalization preferences, and it is marked as "Good" with a label of $\boldsymbol{w}$. IF $score_{\boldsymbol{i}_{k,j,0}} < t$, it means the generated item does not meet the preferences, and it is marked as "Bad" with a label of $\boldsymbol{l}$. Then, we construct positive-negative pairs by comparing fashion items generated within the same outfit. Using the combination formula $C(n,2)$, we can select all possible pairs:
\begin{equation}
\boldsymbol{P}=\left\{\left(\boldsymbol{i}_{k,a,0}, \boldsymbol{i}_{k,b,0}\right) \mid 1 \leq a<b \leq m\right\}.
\end{equation}
We construct the final preference pair set $\boldsymbol{P}_{pref}$ by selecting item pairs where one item matches the preference $w$ and the other does not match the preference $l$:

% \begin{align}
% \boldsymbol{P}_{pref} = 
% \left\{ \left( \boldsymbol{i}_{k,a,0}, \boldsymbol{i}_{k,b,0} \right) \mid \boldsymbol{i}_{k,a,0} = w, \boldsymbol{i}_{k,b,0} = l \right\} 
% & \nonumber \\
% \cup \left\{ \left( \boldsymbol{i}_{k,b,0}, \boldsymbol{i}_{k,a,0} \right) \mid \boldsymbol{i}_{k,a,0} = l, \boldsymbol{i}_{k,b,0} = w \right\}.
% \end{align}

% \begin{align}
% \boldsymbol{P}_{pref} = 
% \left( \boldsymbol{i}_{k,a,0}^w, \boldsymbol{i}_{k,b,0}^l \right) and \left( \boldsymbol{i}_{k,a,0}^l, \boldsymbol{i}_{k,b,0}^w \right).
% \end{align}

\begin{align}
\boldsymbol{P}_{\text{pref}} = 
\left\{ \left( \boldsymbol{i}_{k,a,0}^w, \boldsymbol{i}_{k,b,0}^l \right), \left( \boldsymbol{i}_{k,a,0}^l, \boldsymbol{i}_{k,b,0}^w \right) \right\}.
\end{align}

% {\scriptsize
% \begin{equation}
% \boldsymbol{P}_{pref}=\left\{\left(i_{k,a,0}, i_{k,b,0}\right) \mid i_{k,a,0}=w, i_{k,b,0}=l\right\} \cup\left\{\left(i_{k,b,0}, i_{k,a,0}\right) \mid i_{k,a,0}=l, i_{k,b,0}=w\right\}.
% \end{equation}
% }
This strategy can effectively capture the preferences within the generated fashion items, providing feedback that informs and guides the subsequent model fine-tuning process.

\subsection{\textbf{Model Fine-tuning with Direct Preference Optimization}}

The preference set $\boldsymbol{P}_{pref}$ comprises paired fashion items, denoted as $(\boldsymbol{i}^w, \boldsymbol{i}^l)$, where $\boldsymbol{i}^w$ represents the fashion item that aligns with multiple experts' preferences, and $\boldsymbol{i}^l$ represents the item that does not. Previous work \cite{D3PO} has demonstrated that we can view the inference process as a multi-step Markov Decision Process (MDP).
So we define $\pi(a|s)$ as the policy for taking action $a$ based on state $s$, and view the inference process from timestep $t$ to timestep $t-1$ as taking action $a_t$ from state $s_t$ to state $s_{t-1}$.
In the multi-step inference process of diffusion models, we can obtain the sequence of states and actions:
\begin{equation}
    \sigma_t = \{s_t, a_t, s_{t+1}, a_{t+1}, ..., s_{T}, a_{T}\},  0 \le t \le T.
\end{equation}
Since the predicted fashion items is filled with noise in the early stages of the inference process, it is difficult for both humans and the discriminator model to judge the quality of the images. Therefore, following Reinforcement Learning (RL) ~\cite{RL-1, RL-2} methods, we assume that if the final inference result $\boldsymbol{i}^{w}$ is better than $\boldsymbol{i}^{l}$, then at any timestep during the inference process, the state $\boldsymbol{s}_w$ and action $\boldsymbol{a}_w$ are better than $\boldsymbol{s}_l$ and $\boldsymbol{a}_l$.

\begin{table*}[ht]
\centering
\setlength{\tabcolsep}{3pt}
\caption{Performance comparison between our method and various baselines. "Comp." and "Per." denote compatibility and personalization, respectively. Bold indicates the best results while underline denotes the second best results.}
\vspace{-0.3cm}
\begin{tabular}{lccccccc|ccccccc}
\toprule
Dataset & \multicolumn{7}{c}{iFashion} & \multicolumn{7}{c}{Polyvore-U} \\
\cmidrule(lr){2-8} \cmidrule(lr){9-15} 
Task & \multicolumn{4}{c}{PFITB} & \multicolumn{3}{c}{GOR} &\multicolumn{4}{c}{PFITB} & \multicolumn{3}{c}{GOR} \\
\cmidrule(lr){2-5} \cmidrule(lr){6-8} \cmidrule(lr){9-12} \cmidrule(lr){13-15} 
 Evaluation metric & IS & IS-acc & Comp. & Per. & IS & IS-acc & Per. & IS & IS-acc & Comp. & Per. & IS & IS-acc & Per.\\
\midrule
% OutfitGAN & 202.60 & 9.52 & 0.19 & 18.07 & 16.88 & 0.63 & 0.03 & 21.48 & - & - & - & - & - & - & - \\
SD-v1.5~\cite{SD}* & 22.54 & 0.76 & 0.08 & 46.31 & 23.20 & 0.77 & 46.45 & 17.10  & 0.73  & 0.70  & 51.05  & 16.95  & 0.73 & 50.99\\
SD-v2*~\cite{SD} & 21.66 & 0.71 &  0.04 & 46.60 & 22.19 & 0.74  & 46.60 & 14.83  & 0.68  & 0.60  & 51.29  & 14.88  & 0.67 & 51.23\\
SD-v1.5~\cite{SD} & 26.76 & 0.83 & 0.46 & 53.16 & 26.90 & 0.84  & 53.24 & 17.12  & 0.72  & 0.75  &  58.20 &  17.24  &  0.72 &  58.16\\
SD-v2~\cite{SD} & 25.85 & 0.80 & 0.39 & 52.99 & 25.82 & 0.82 & 53.06 & 15.59  & 0.67  & 0.71  & 58.79  & 16.33  & 0.70 & 58.91\\
SD-naive~\cite{SD} & 25.45 & 0.80 & 0.36 & 52.95 & 25.43 & 0.81 & 52.95 & 15.45  & 0.66  & 0.73  & 59.24  & 15.48  &  0.67 & 59.12\\
ControlNet~\cite{controlNet} & 27.76 & 0.81 & 0.16 & 49.90 & 28.49 & 0.82 & 49.91 & 18.93   & 0.77  & 0.73  & 55.44  & \underline{19.21}  & 0.77 & 55.40\\
DiFashion~\cite{DiFashion} & \underline{29.99} & \underline{0.90} & \underline{0.58} & \underline{55.86} & \underline{30.04} & \underline{0.90} & \underline{55.54} & \underline{19.67}  & \underline{0.84}  & \underline{0.80}  & \underline{61.44}  & 18.95  & \underline{0.83} & \underline{61.16}\\
\midrule
\rowcolor{gray!30} 
\textbf{FashionDPO(Ours)} & \textbf{33.80} &  \textbf{0.91} & \textbf{0.74} & \textbf{60.39} & \textbf{32.37} & \textbf{0.91} & \textbf{59.98} & \textbf{24.14} & \textbf{0.89} & \textbf{0.83} & \textbf{64.67} & \textbf{24.93} & \textbf{0.87} & \textbf{64.79}\\
\bottomrule
\end{tabular}
\vspace{-0.3cm}
\label{tab:quantitative}
\end{table*}

In Section \ref{inference}, we saved the latent image variables from $T$ timesteps of each generated fashion item $\boldsymbol{i}$ as the state $s$. Based on the positive-negative pairs $(\boldsymbol{i}^w, \boldsymbol{i}^l)$ obtained from the feedback, we have the states $\boldsymbol{s}^w = \{\boldsymbol{i}_0^w, ..., \boldsymbol{i}_{T}^w\}$ and $\boldsymbol{s}^l = \{\boldsymbol{i}_0^l, ..., \boldsymbol{i}_{T}^l\}$. As shown in Figure \ref{fig:model}, at timestep $t \in \{1,...,T\}$, saved latent $\boldsymbol{i}_t^{w}$ undergoes noise prediction, guided by the conditions $(\boldsymbol{m}_k, \boldsymbol{h}_k, \boldsymbol{p}_k)$, where the trainable and the frozen UNet respectively obtain $\boldsymbol{\epsilon}_{t,\theta}^w$ and $\boldsymbol{\epsilon}_{t,ref}^w$. Then we estimate the original noise-free latent variable $\boldsymbol{\hat{i}}_0^w$ using the noise $\boldsymbol{\epsilon}_{t,\theta}^w$  and the current latent variable  $\boldsymbol{i}_t^w$:
\begin{equation}
\boldsymbol{\hat{i}}_0^w=\frac{\boldsymbol{i}_t^w-\sqrt{1-\alpha_t} \cdot \epsilon_{t,\theta}^w\left(\boldsymbol{i}_t^w, t, \boldsymbol{c}\right)}{\sqrt{\alpha_t}},
\end{equation}
where $\alpha_t$ is a decay coefficient that we pre-calculate and store during the inference process, and $\boldsymbol{c}$ represents multiple conditions. Then we can calculate the mean $\boldsymbol{\mu}_{t-1}$ and variance $\sigma_t^2$ of the noise distribution at time step $t-1$:
\begin{equation}
\begin{gathered}
\boldsymbol{\mu}_{t-1}=\sqrt{\alpha_{t-1}} \boldsymbol{\hat{i}}_0^w+\sqrt{1-\alpha_{t-1}} \epsilon_{t,\theta}^w\left(\boldsymbol{i}_t^w, t, \boldsymbol{c}\right), \\
\sigma_t^2=\frac{1-\alpha_{t-1}}{1-\alpha_t} \cdot\left(1-\frac{\alpha_t}{\alpha_{t-1}}\right).
\end{gathered}
\end{equation}
Based on the above parameters, the latent variable $\boldsymbol{i}_{t-1}^w$ at timestep $t-1$ follows a Gaussian distribution:
\begin{equation}
\begin{gathered}
\pi_{\mathrm{\theta}}\left(\boldsymbol{i}_{t-1}^w \mid \boldsymbol{i}_t^w, \boldsymbol{c}\right) \sim \mathcal{N}\left(\boldsymbol{\mu}_{t-1}, \sigma_t^2\right), \\
\pi_{\mathrm{\theta}}\left(\boldsymbol{i}_{t-1}^w \mid \boldsymbol{i}_t^w, \boldsymbol{c}\right)=\frac{1}{\sqrt{2 \pi \sigma_t^2}} \exp \left(-\frac{\left(\boldsymbol{i}_{t-1}^w-\boldsymbol{\mu}_{t-1}\right)^2}{2 \sigma_t^2}\right),
\end{gathered}
\end{equation}
where $\boldsymbol{i}_{t-1}^w$ refers to the latent variable that we save during the inference process at the corresponding timestep $t-1$. The same approach can be applied to obtain other parametrized policies $\pi_\theta$ and $\pi_{ref}$. Then, based on the use of Eq.\ref{dpoloss}, we derive the loss function of the fashionDPO model:

{\footnotesize 
\begin{equation}
\mathcal{L}_{\mathrm{DPO}}=-\mathbb{E}_{\left(s^w,s^l \right)}\left[\log \sigma\left(\beta_w \log \frac{\pi_\theta\left(\boldsymbol{i}_{t-1}^w \mid \boldsymbol{i}_t^w, \boldsymbol{c}\right)}{\pi_{\mathrm{ref}}\left(\boldsymbol{i}_{t-1}^w \mid \boldsymbol{i}_t^w, \boldsymbol{c}\right)}-\beta_l \log \frac{\pi_\theta\left(\boldsymbol{i}_{t-1}^l \mid \boldsymbol{i}_t^l, \boldsymbol{c}\right)}{\pi_{\mathrm{ref}}\left(\boldsymbol{i}_{t-1}^l \mid \boldsymbol{i}_t^l, \boldsymbol{c}\right)}\right)\right],
\end{equation}
}
where $\beta_w$ and $\beta_l$ denote parameters used to control preference and non-preference biases, $\sigma(\cdot)$ denotes the sigmoid function. For each preference pair $(\boldsymbol{i}^w, \boldsymbol{i}^l)$, fine-tuning occurs at each timestep $t \in \{T,T-1,...,1\}$, repeated $T-1$ times. Subsequently, the next preference pair is drawn from the preference dataset $\boldsymbol{P}_{pref}$, continuing this process until all preference data has been used for fine-tuning.

% And we leverage an analytical mapping from reward functions to optimal policies, transform a loss function over reward functions into a loss function over policies. 

% 由于loss由偏好结果给出，而非生成结果与groundtruth之间的差异，这有效地解决了groundtruth在训练中对搭配但与groundtruth不一致的fashion items错误惩罚的问题。
% Problem:imited diversity, misalignment between tasks and objectives
% Challenge:%The challenge is the design of AI evaluation models and the mechanisms for updating feedback.

\section{Experiments}
We conduct a series of experiments on two established datasets of iFashion and Polyvore-U to evaluate our method, including both quantitative and qualitative analysis, as well as intricate model study. The experiments are designed to answer following questions:
\begin{itemize}
\item \textbf{RQ1:} The effectiveness of the fine-tuning framework. After multiple rounds of feedback-based fine-tuning iterations, does the model show improvements in quantitative metrics compared to DiFashion and other generative models?
\item \textbf{RQ2:} During multiple iterations, how does FashionDPO perform in terms of image diversity and personalization compared to itself and other baselines?
\item \textbf{RQ3:} How do data and time costs, alternative implementations of experts, and hyper-parameter adjustments affect the performance of the FashionDPO framework?
\end{itemize}

%In the quantitative results, we evaluate the FashionDPO using multi-perspective metrics. In the qualitative results, we invite professional fashion designers to conduct expert-level human evaluations. 
%In the model study, we further explore the alternative implementations of experts and other parameter settings.

% we perform an analysis of time complexity and fine-tuning stability, and explore the capability of FashionDPO in improving the model's learning of fashion compatiby principles.
% incorporating pairing knowledge from multimodal large models to further evaluate the fine-tuning effect. 
% By analyzing these metrics, we validate the effectiveness of the FashionDPO fine-tuning framework. 

\begin{figure*}
  \centering
  \setlength{\abovecaptionskip}{0.1cm} 
  \setlength{\belowcaptionskip}{-0.3cm} 
  \includegraphics[width=1.0\textwidth]{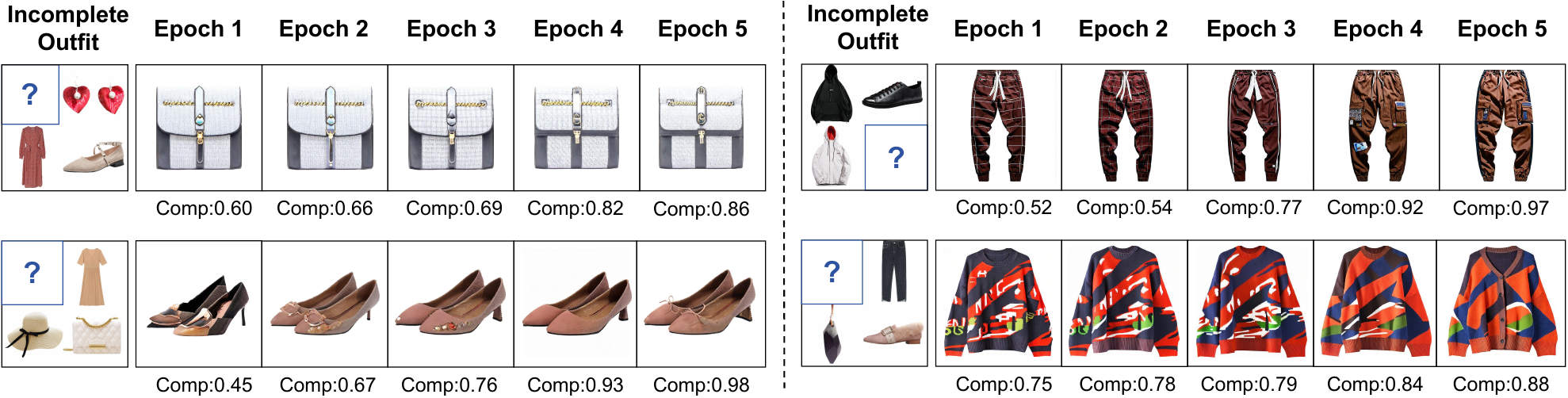} 
  \caption{
   Epoch-wise comparison of FashionDPO's performance across different fine-tuning epochs. As epochs increase, the compatibility metric indicates that the generated fashion items better match the incomplete outfit.}
  \label{fig:result_epochs}
\end{figure*}

\subsection{Experimental Settings}

% \subsubsection{\textbf{Datasets}}
% We employ two datasets: \ie iFashion~\cite{POG} and polyvore-U~\cite{Polyvore}, which include fashion outfit and user-fashion item interaction history. We preprocessed the two datasets separately for use in the current task.
% For the iFashion dataset, we select 50 common fashion categories from the dataset and filter the fashion products accordingly. Each outfit consists of four fashion products. The filtered dataset includes the categories and image information of the fashion products, as well as user interaction history with outfits. For the Polyvore-U dataset, since it only contains basic category information such as top and bottom, we used the classifier Inception-V3 ~\cite{Inception-V3} fine-tuned on the iFashion dataset to perform image recognition and classification on the fashion products. We divided the dataset into training and testing sets, ensuring that each user has an interaction history with more than 5 outfits. The model is fine-tuned based on the feedbacks from the sampling results of training set and calculate evaluation metrics based on testing set.

\subsubsection{\textbf{Baselines}}
For the two tasks of PFITB and GOR, we compare our model with the following baselines:
\textbf{1) SD-v1.5}~\cite{SD}: It's a latent space diffusion model. In the model names, with "*" indicates a pre-trained model, while without "*" indicates that the model has been fine-tuned on the fashion dataset.
\textbf{2) SD-v2}: It's an upgraded version of SD-v1.5. The same naming convention.
\textbf{3) SD-naive}: It's a fine-tuned model based on SD-v2, where concatenate mutual influence and history condition as condition.
\textbf{4) ControlNet}~\cite{controlNet}: It's an extension model based on SD, which controls the details of generated images by introducing additional conditional inputs.
\textbf{5) DiFashion}~\cite{DiFashion}: It's the SOTA generative recommendation model based on SD-v2, which uses Classifier-Free Guidance (CFG)~\cite{CFG} to tightly align the control conditions of the generated fashion images.

\subsubsection{\textbf{Datasets}}
We follow the previous works~\cite{DiFashion,PFOG} and use the datasets of iFashion~\cite{POG} and Polyvore-U~\cite{polyvore_u}, which include the required data of both fashion outfit and user-fashion item interactions.
For the iFashion dataset, we select 50 common fashion categories from the dataset and filter the fashion products accordingly. Each outfit consists of four fashion products. For the Polyvore-U dataset, since it only contains basic category information such as top and bottom, we used the classifier Inception-V3 ~\cite{Inception-V3} fine-tuned on the iFashion dataset to perform image recognition and classification on the fashion products. We divided the dataset into training and testing sets, ensuring that each user has an interaction history with more than five outfits. The model is fine-tuned based on the feedbacks from the sampling results of training set and calculate evaluation metrics based on testing set.
% The filtered dataset includes the categories and image information of the fashion products, as well as user interaction history with outfits.

\begin{figure}
  \centering
  \setlength{\abovecaptionskip}{0.1cm} 
  \setlength{\belowcaptionskip}{-0.5cm} 
  \includegraphics[width=0.45\textwidth]{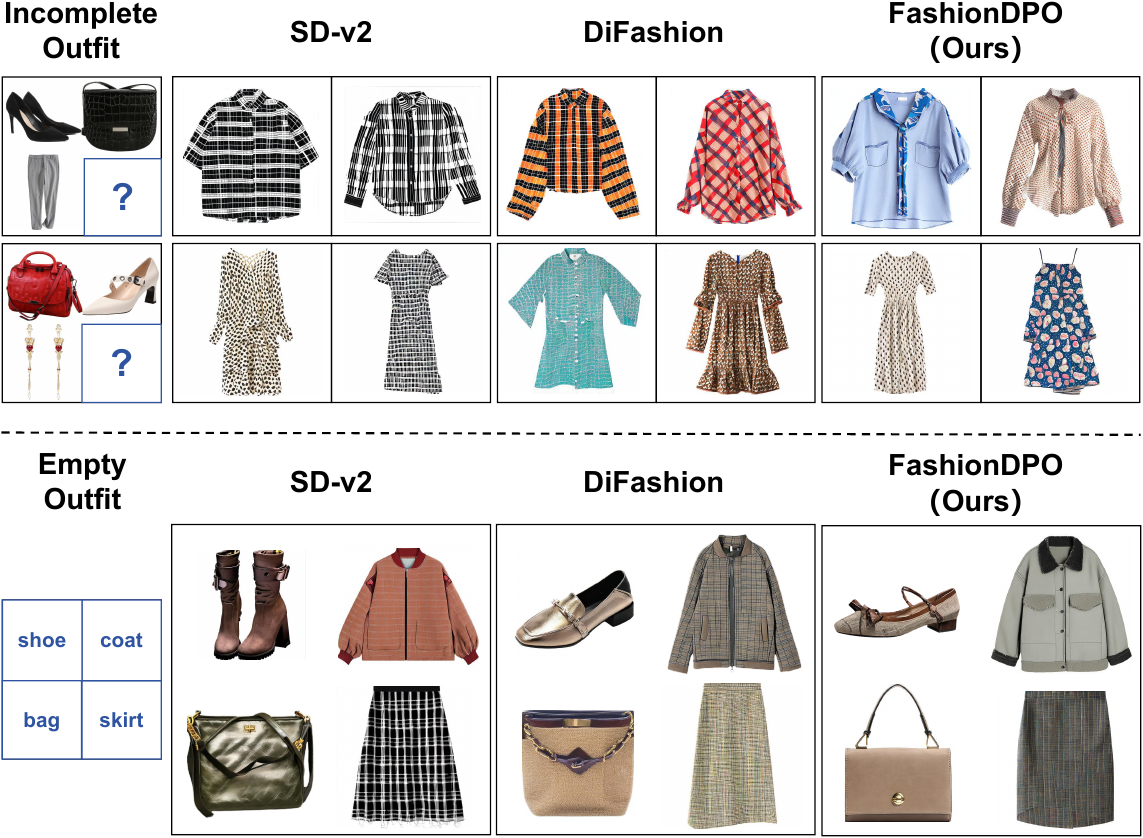} 
  \caption{
   Model-wise comparison of different models' generative capabilities: PFITB task above the line, GOR task below.
   }
  \label{fig:model_study_vbpr}
\end{figure}

\subsubsection{\textbf{Evaluation Metrics}}
We employ a list of quantitative evaluation metrics, covering three major evaluating perspectives: 
\textbf{1) Quality}: We use Inception Score (IS) to evaluate the quality of the generated images. Additionally, we use IS-Accuracy (IS-acc) to further assess whether the generated images are correct regarding its category-level semantics. 
\textbf{2) Compatibility (Comp.)}: We follow the previous work~\cite{DiFashion} and use the discriminator in OutfitGAN~\cite{OutfitGAN} to calculate the compatibility of the generated outfit in PFITB task.
\textbf{3) Personalization (Per.)}: We use the foundation model CLIP~\cite{CLIP} to extract the image embeddings of the items that a user has interacted in the history. Then we calculate the cosine similarity between generated fashion items and history image embeddings.

\subsubsection{\textbf{Implementation Details}}
% In the Fashion Image Generation without Feedback module, if it is a PFITB task, mutual information $\{\boldsymbol{i}_{k,t}\}_{k=2}^n$ is obtained by adding the corresponding timestep $t$ of noise to the incomplete outfit. If it is a GOR task, $\{\boldsymbol{i}_{k,t}\}_{k=2}^n$ is obtained through denoising $T-t$ steps random noise. 
% We fine-tune the model on the PFITB task, and the fine-tuned model can also be used for the GOR task.

In our experiments, we load the pre-trained DiFashion model to initialize the model parameters.
The training subset consists of 1,000 outfits randomly selected from the iFashion or Polyvore-U dataset.
During the sampling phase, for each outfit, we generate $j=7$ fashion items as $\boldsymbol{I}_0$, with time step set to 50. 
At each time step, we save the latent variables and the noise predicted by diffusion model as data for fine-tuning.
And in the Feedback Generation from Multiple Experts module, we set $\alpha_q=\alpha_c=\alpha_p=1$, and threshold $t$ is the average of $score$.
In the Model Fine-tuning with Direct Preference Optimization module, we perform LoRA~\cite{lora} fine-tuning on the 50 saved timesteps.
% , adding low-rank matrices $\boldsymbol{A}$ and $\boldsymbol{B}$ to the attention layer of U-Net
At each time step, we compute the loss $\mathcal{L}_{\mathrm{DPO}}$ and update the gradients.
Repeat this process until the preferences in each outfit have been learned by the model.
%, where $\boldsymbol{B}$ is initialized with zeros. 
%The updated weight matrix $\boldsymbol{W}$ after LoRA adaptation is $\boldsymbol{W} = \boldsymbol{W} + \boldsymbol{AB}$. 
In the next epoch, a new training subset of 1000 outfits is selected, and repeat the "sampling - feedback - fine-tuning" process. In our experiment, we fine-tuned DiFashion for five epochs to obtain the final FashionDPO model.
% We set the learning rate to $1e-5$ and fine-tune five epochs.
Meanwhile, we conduct parameter search and the final model uses $\beta_w = \beta_l = 0.5$.

% t \in \{1,2,...,T\}
% lora
% T,size,等dpo参数设置
% 只在pfitb任务上面进行微调，微调之后的模型可以直接用于gor任务

\begin{table}
\centering
\setlength{\tabcolsep}{0.5pt}
\caption{Quantitative results on different ablated models, where bold font indicates the best results while underline denotes the second best results.}
\vspace{-0.3cm}
\begin{tabular}{lccccccc}
\toprule 
iFashion & \multicolumn{4}{c}{PFITB} & \multicolumn{3}{c}{GOR}  \\
\cmidrule(lr){2-5} \cmidrule(lr){6-8} 
Method & IS & IS-acc & Comp. & Per. & IS & IS-acc & Per. \\
\midrule
% OutfitGAN & 202.60 & 9.52 & 0.19 & 18.07 & 16.88 & 0.63 & 0.03 & 21.48 & - & - & - & - & - & - & - \\
w/o Feedback & 29.46 & 0.88 & 0.59 & 55.43 & 29.30 & 0.89 & 55.89 \\
w/o Quality Expert & 30.14 & 0.88 & \underline{0.70} & \underline{59.82} & 29.94 & 0.90 & \underline{58.41}\\
w/o Comp. Expert & 32.79 & \underline{0.90} & 0.61 & 58.27 & 31.90 & 0.90 & 57.80 \\
w/o Per. Expert & \underline{33.10} & \underline{0.90} & 0.68 & 56.92 & \underline{32.41} & \underline{0.91} & 56.25\\
\textbf{FashionDPO} & \textbf{33.80} &  \textbf{0.91} & \textbf{0.74} & \textbf{60.39} & \textbf{32.37} & \textbf{0.91} & \textbf{59.98} \\
\bottomrule
\end{tabular}
\vspace{-0.4cm}
\label{tab:ablation}
\end{table}

% \begin{table}
% \centering
% \setlength{\tabcolsep}{0.5pt}
% \caption{Quantitative results on different ablated models, where bold font indicates the best results while underline denotes the second best results.}
% \vspace{-0.3cm}
% \begin{tabular}{lccccccc}
% \toprule 
% \#iFashion & \multicolumn{4}{c}{PFITB} & \multicolumn{3}{c}{GOR}  \\
% \cmidrule(lr){2-5} \cmidrule(lr){6-8} 
% Method & IS$\uparrow$ & IS-acc$\uparrow$ & Comp.$\uparrow$ & Per.$\uparrow$ & IS$\uparrow$ & IS-acc$\uparrow$ & Per.$\uparrow$ \\
% \midrule
% % OutfitGAN & 202.60 & 9.52 & 0.19 & 18.07 & 16.88 & 0.63 & 0.03 & 21.48 & - & - & - & - & - & - & - \\
% w/o Feedback & 29.46 & 0.88 & 0.59 & 55.43 & 29.30 & 0.89 & 55.89 \\
% w/o Quality Expert & 30.14 & 0.88 & \underline{0.70} & \underline{59.82} & 29.94 & 0.90 & \underline{58.41}\\
% w/o Comp. Expert & 32.79 & \underline{0.90} & 0.61 & 58.27 & 31.90 & 0.90 & 57.80 \\
% w/o Per. Expert & \underline{33.10} & \underline{0.90} & 0.68 & 56.92 & \underline{32.41} & \underline{0.91} & 56.25\\
% \textbf{FashionDPO} & \textbf{33.80} &  \textbf{0.91} & \textbf{0.74} & \textbf{60.39} & \textbf{32.37} & \textbf{0.91} & \textbf{59.98} \\
% \bottomrule
% \end{tabular}
% \vspace{-0.3cm}
% \label{tab:ablation}
% \end{table}

\subsection{Quantitative Results (RQ1)}
In this section, we compare FashionDPO with baselines across multiple metrics and validate the effectiveness of feedback and multiple experts through ablation experiments.

\subsubsection{\textbf{Overall Performance Comparison}}
We aim to evaluate the model's performance on different datasets and tasks.
The results are shown in Table \ref{tab:quantitative}, and we analyze them as follows:
% The quantitative evaluation results are presented in Table \ref{tab:quantitative}. We will analyze the quantitative results through comparisons across tasks and models:

% \textbf{1) Datasets}: Our Fashion model performs better in the image quality metrics IS and IS-acc on the iFashion dataset. On the Polyvore-U dataset, Fashion achieves higher scores in the compatibility and personalization metrics. 

\begin{figure*}[ht]
    \centering
    \setlength{\abovecaptionskip}{0.1cm} 
    \setlength{\belowcaptionskip}{-0.1cm} 
    \includegraphics[width=0.95\textwidth]{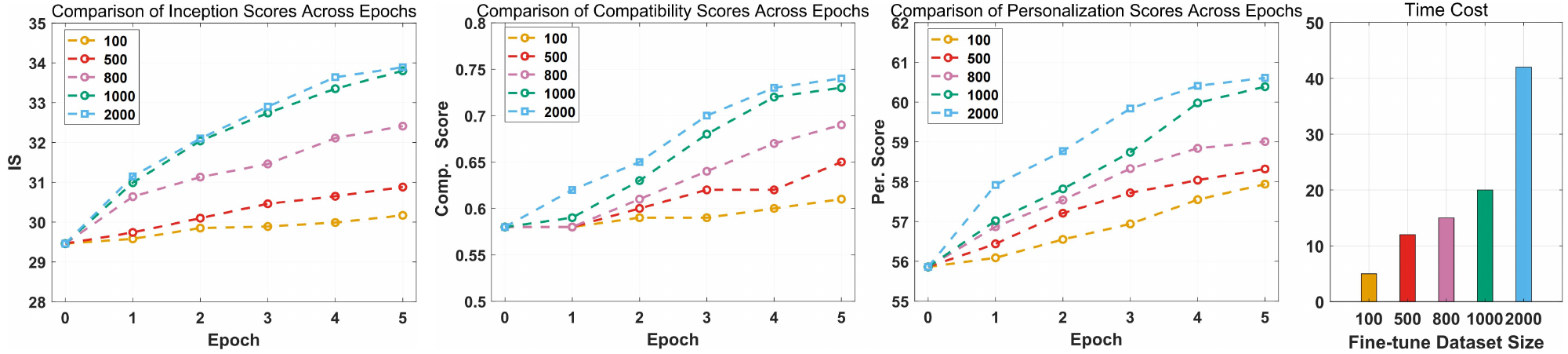} 
    \caption{We fine-tune our model on a subset with $n$ outfits, where $n \in \{100, 500, 800, 1000, 2000\}$, to explore the impact of varying datase size on model performance. Lines represent models fine-tuned on different subsets, with the x-axis as epochs and the y-axis as the evaluation metric. Bars show the time cost at per epoch (inference, feedback, fine-tuning) for different subsets.}
    \label{multi_sizes}
\end{figure*}

% \begin{figure*}[ht]
%     \centering
%     \setlength{\abovecaptionskip}{0.2cm} 
%     \setlength{\belowcaptionskip}{-0.1cm} 
%     \subfloat{
%     \label{multi_size_is}
%     \includegraphics[width=0.245\linewidth]{images/multi_size_IS.pdf}
%     }
%     \hspace{-0.2cm}
%     \subfloat{
%     \label{multi_size_com}
%     \includegraphics[width=0.26\linewidth]{images/multi_size_com.pdf}
%     }
%     \hspace{-0.3cm}
%     \subfloat{
%     \label{multi_size_per}
%     \includegraphics[width=0.26\linewidth]{images/multi_size_per.pdf}
%     }
%     \hspace{-0.2cm}
%     \subfloat{
%     \label{multi_size_timecost}
%     \includegraphics[width=0.17\linewidth]{images/model_study_timecost.pdf}
%     }
%     \caption[]{We fine-tune our model on a subset with $n$ outfits, where $n \in \{100, 500, 800, 1000, 2000\}$, to explore the impact of varying datase size on model performance. Lines represent models fine-tuned on different subsets, with the x-axis as epochs and the y-axis as the evaluation metric. Bars show the time cost at per epoch (inference, feedback, fine-tuning) for different subsets.}
%     \vspace{-0.2cm}
%     \label{multi_sizes}
% \end{figure*}

\textbf{1) Tasks}: Our FashionDPO achieves similar scores across both tasks, indicating that the model fine-tuned on the PFITB task can smoothly transition to the GOR task. This demonstrates the generalizability of the fine-tuning framework.

\textbf{2) Models}: The results show that our FashionDPO achieves advantages over other baselines across most metrics. For the IS and IS-acc metric, the improvement of our model is significantt, indicating that our model generates more diverse results. In comparison, other baselines, such as SD-native and ControlNet, use supervised learning methods for fine-tuning, resulting in limited improvement. This indicates that our model has learned broader styling techniques and design principles.
Based on the Comp. and Per. metrics, our FashionDPO demonstrates improvements over the SOTA method DiFashion. This indicates that our fine-tuning framework can effectively incorporate the knowledge of different fashion experts into the model through feedback, enabling it to learn diverse knowledge from the fashion domain.

% since our FashionDPO is fine-tuned based on feedback from multiple experts, we can avoid incorrect penalization of generated results that are well-matched but differ from the ground truth in image features. In comparison, other baselines, such as SD-native and ControlNet, use supervised learning methods for fine-tuning, resulting in limited improvement in the IS metric. From the Comp. and Per. metrics, it can be observed that our FashionDPO shows improvement compared to DiFashion after multiple epochs of iteration. This indicates the effectiveness of our fine-tuning framework, which incorporates compatibility and personalization experts for evaluation. Through feedback, the model learns specialized fashion domain knowledge, thereby enhancing its outfit compatibility and personalization capabilities.

\subsubsection{\textbf{Ablation Study}}
We perform ablation experiments on the iFashion dataset, with each experiment fine-tuned for 5 epochs.

\noindent \textbf{w/o Feedback.}
To verify the effectiveness of multiple experts' feedback, we removed it and fine-tuned the pre-trained DiFashion model on the iFashion dataset using LoRA.

% This framework conducts a multiple experts' evaluation on the sampling results, considering three aspects: quality, compatibility, and personalization. As a comparison, we assign random scores to the generated results to observe whether randomly selecting positive and negative samples will lead to a decline in model performance.

\noindent \textbf{w/o Multiple Experts.}
To verify the necessity of fashion experts, we conducted experiments by removing one of the three experts (\ie quality, compatibility, and personalization expert) .

The results are shown in Table \ref{tab:ablation}, without feedback from multiple experts, the improvements across various metrics are minimal. By removing one of the three experts, it can be observed that without the Quality Expert has minimal impact on compatibility and personalization, while significantly affecting the IS metric. It can also be observed that removing either the Compatibility Expert or the Personalization Expert leads to significant declines in their corresponding evaluation metrics.
This demonstrates that our three designed evaluation perspectives are comprehensive and objective, as the absence of any single one leads to a decline in model performance.

\subsection{Qualitative Results (RQ2)}
%In this section, we present some examples of fashion items created by FashionDPO. And we invite professional fashion designers to conduct expert-level human evaluations.
In addition to quantitative results, we present qualitative analysis from multiple perspectives. To verify the objectivity and effectiveness of the fine-tuning results, we invite professional fashion designers to conduct expert-level human evaluations.

\begin{table}
  \centering 
  \setlength{\tabcolsep}{2.pt}
  \caption{Results of expert-level human evaluation.}
  \vspace{-0.3cm}
  \begin{tabular}{l|cccc}
  \toprule Model & D1-Style & D2-Color & D3-Fabric & D4-Variety \\
  \midrule DiFashion & 2.73±0.63 & 2.74±0.64 & 2.97±0.67 & 2.91±0.47\\
  \textbf{FashionDPO(Ours)} & 4.08±0.52 & 3.87±0.37 & 3.63±0.50 & 3.22±0.43\\
  \bottomrule
  \end{tabular}
  \label{tab:human_evaluation}
  \vspace{-0.4cm}
\end{table}

\subsubsection{\textbf{Epoch-wise Comparison}}
To validate that the model can learn fashion domain knowledge throughout the fine-tuning epochs, we present the model's performance at different epochs.
As shown in Figure \ref{fig:result_epochs}, with the increase in the number of epochs, the fashion items generated by our FashionDPO become progressively more compatible with the incomplete outfit. 
For example, in Epochs 1 and 2, although the generated pants and shoes roughly match the style, there are inconsistencies in materials or color tones. By Epochs 4 and 5, the generated pants and shoes align more closely with the theme of the complete outfit, and their materials and colors gradually become more consistent with the existing clothing.
Notably, to facilitate the comparison of model performance across different epochs, we started from the same Gaussian noise image and used the DDIM~\cite{DDIM} scheduler for sampling. And if we use the DDPM~\cite{ddpm} or PNDM~\cite{pndm} scheduler or initialize with different noise, the results will be diverse.
%Since the DDIM sampling process is deterministic, the generated results remain consistent given the same parameters and initial noise. As the model parameters are updated during fine-tuning, the results in the figure appear similar yet different textures and details in the results indicate that the model has learned fashion domain knowledge throughout fine-tuning epochs.

\subsubsection{\textbf{Model-wise Comparison}}
In order to qualitative demonstrate the capability difference across various models, we present a case study by comparing the generated results of three models on two tasks.
As shown in Figure \ref{fig:model_study_vbpr}, the results generated by our FashionDPO are more diverse. In the first row of the PFITB task results, the incomplete outfit includes a pair of black high heels and a black bag. The SD-v2 and DiFashion generate plaid shirts, which have a casual style and appear mismatched. Our FashionDPO generates light blue shirt and beige cardigan align better with the given fashion items. In the second row, the dresses generated by FashionDPO exhibit variations in style (such as floral patterns), making them more visually appealing. And in the GOR task results, FashionDPO generates a light-colored bag and brown shoes that display strong cohesion with the gray coat and skirt. While the outfits generated by SD-v2 and DiFashion are reasonable, their overall style appears somewhat inconsistent.
By comparing with other baselines, it can be seen that the fashion products generated by FashionDPO adhere to fashion design principles and better meet the personalized needs of users.

\begin{figure}[t]
    \centering
    \setlength{\abovecaptionskip}{0.1cm} 
    \setlength{\belowcaptionskip}{-0.3cm} 
    \includegraphics[width=0.45\textwidth]{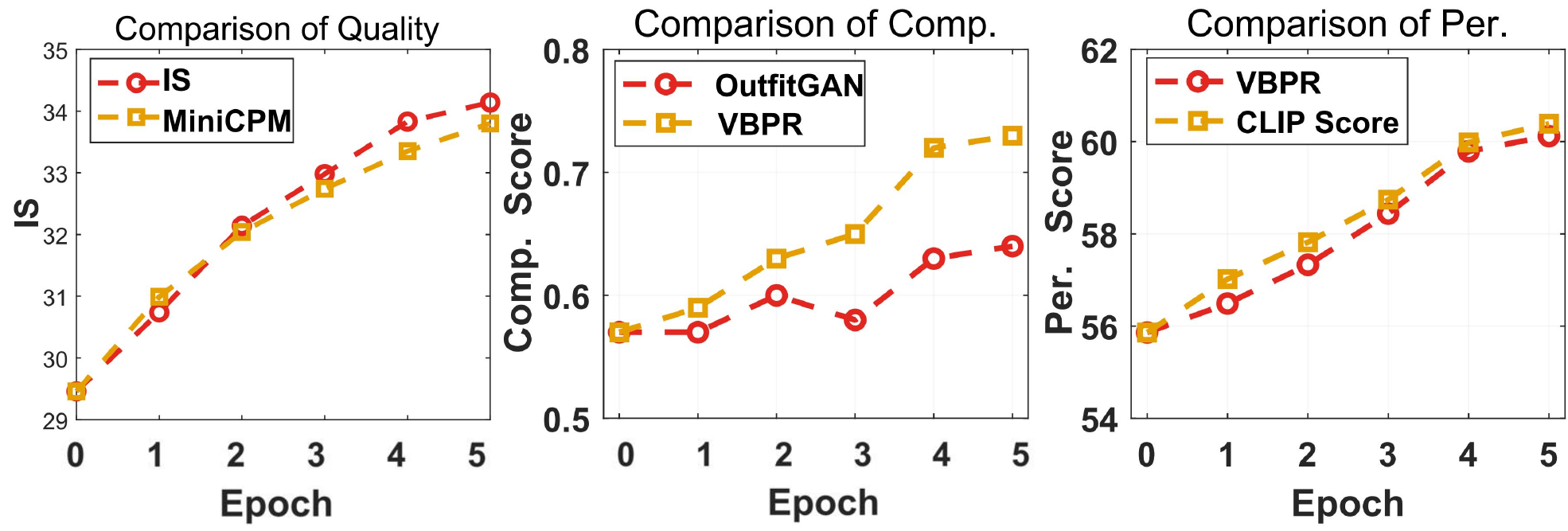} 
    \caption{The impact of expert implementations. Each line corresponds to one type of expert implementation, with the x-axis as epochs and the y-axis as the evaluation metric.
    }
    \label{multi_experts}
\end{figure}

% \begin{figure}[t]
%     \centering
%     % \setlength{\abovecaptionskip}{0.1cm} 
%     % \setlength{\belowcaptionskip}{-0.3cm} 
%     \subfloat{
%     \label{multi_qua}
%     \includegraphics[width=0.31\linewidth]{images/model_study_multi_qua.pdf}
%     }
%     \hspace{-0.15cm}
%     \subfloat{
%     \label{multi_com}
%     \includegraphics[width=0.32\linewidth]{images/model_study_multi_com.pdf}
%     }
%     \hspace{-0.15cm}
%     \subfloat{
%     \label{multi_per}
%     \includegraphics[width=0.32\linewidth]{images/model_study_multi_per.pdf}
%     }
%     \caption[]{The impact of expert implementations. Each line corresponds to one type of expert implementation, with the x-axis as epochs and the y-axis as the evaluation metric.
%     }
%     \label{multi_experts}
% \end{figure}

% by comparing with other models, it can be seen that the fashion products generated by FashionDPO adhere to fashion design principles and better meet the personalized needs of users.

% We present IS, compatibility, personalization, and time cost along with the epoch growing.

\subsubsection{\textbf{Human Evaluation by Fashion Experts}}
We invite five professional fashion designers \footnote{From Jiangnan University, School of Digital Technology \& Innovation Design, 214122, Wuxi, China} aged 18-30 to conduct expert-level double-blind human evaluations. We use a five-level scoring protocol to score 30 sets of results generated by two different models regarding to four evaluating aspects of D1-D4 (style, color, fabric, variety).
The scores are discretized to five levels: very satisfied, satisfied, average, dissatisfied, and very dissatisfied, corresponding to 5 to 1 point, respectively. 
Finally, we collect valid scoring results from five designers for analysis. 
% For more details of the human evaluation by fashion experts, please refer to Appendix \ref{sec:user_study}.

As shown in Table \ref{tab:human_evaluation}, it can be seen that our FashionDPO is recognized by the majority of fashion experts, achieving higher mean scores with lower variance.
Furthermore, we conduct interviews with the participating fashion designers and gathered some common subjective feedback. For instance, almost all fashion experts indicated that FashionDPO demonstrate better diversity and outfit compatibility compared to DiFashion, especially in the matching of formal wear and evening dress styles. In addition to discussions on style, the performance of individual items also catch the attention of fashion designers. For example, FashionDPO deliver satisfactory generation results in categories such as bags, shoes and dresses.

% \begin{figure*}
%   \centering
%   \includegraphics[width=0.95\textwidth]{images/model_study_vbpr_outfitgan.pdf} 
%   \caption{
%    Model Study - OutfitGAN -- VBPR
%   }
%   \label{fig:model_study_vbpr}
% \end{figure*}

\subsection{Model Study (RQ3)}
We further study several essential properties of our model.

\subsubsection{\textbf{Data and Time Cost Analysis}}
We test the data and time required for fine-tuning. 
During iterative fine-tuning, we evaluate the model's performance differences across different epochs using three major evaluation perspectives: Quality - IS, Compatibility - Comp. Score, and Personalization - Per. Score.
As shown in Figure \ref{multi_sizes}, we present the results fine-tuned on five subsets of different sizes. 
As the size of the subset increases, the model can learn fashion knowledge within less epochs, while the time required for fine-tuning still increases. Notably, when the number of epochs is set to five, the performance of the subset with 1,000 is similar to that with 2,000. However, the time doubles for the subset with 2,000 outfits. Therefore, we choose 1,000 outfits as the default  fine-tuning subset.

% and experiments have demonstrated that FashionDPO can effectively enhance the model's generative recommendation capability with fine-tuning on a small-scale sample, making it suitable for application in tasks.

% We assign a certain weight to the feedback from multiple experts separately, and the final score is a weighted result of three aspects (\ie quality, compatibility, and personalization).

\begin{table}
\centering
\setlength{\tabcolsep}{0.2pt}
\caption{The performance of different evaluation models on the test set. The columns of the table represent three evaluation metrics, while the rows represent the evaluation experts corresponding to different perspectives.}
\vspace{-0.2cm}
\small
\begin{tabular}{l|cccccc}
\toprule 
\multirow{2.2}{*}{Metric}  & \multicolumn{2}{c}{Quality} & \multicolumn{2}{c}{Compatibility} & \multicolumn{2}{c}{Personalization} \\
\cmidrule(lr){2-3} \cmidrule(lr){4-5} \cmidrule(lr){6-7} 
& \textbf{MiniCPM} & GPT-4  &  OutfitGAN & \textbf{VBPR}  &  \textbf{CLIP Score} & VBPR \\
\midrule
Accuracy & 89.10\% & 82.50\% & 70.10\% & 84.80\% & 87.40\% & 86.60\% \\
Mean & 0.56±0.15 & 0.64±0.21 & 0.47±0.14 & 0.44±0.05 & 0.53±0.04 & 0.56±0.05 \\
Pair-t & \multicolumn{2}{c}{$9.29\times10^{-4}$} &  \multicolumn{2}{c}{$6.52\times10^{-6}$}  &  \multicolumn{2}{c}{$8.31\times10^{-3}$} \\
\bottomrule
\end{tabular}
\vspace{-0.3cm}
\label{tab:test_multi_expert}
\end{table}

\subsubsection{\textbf{Alternative Implementations of Experts}} \label{expert implementation}
To validate whether our method is sufficiently generalizable for expert implementations, we replace the expert model in each of the three evaluation perspectives and fine-tune for five epochs on the same data subset.
% The multi-expert feedback module includes three experts covering diverse evaluation perspectives: quality, compatibility and personalization.
% We explore the impact of feedback from experts with different capabilities. 
The results are shown in Figure \ref{multi_experts}, where we present the performance for different expert implementations, with the x-axis as epochs and the y-axis as the evaluation metric. 
We can see that replacing the quality and personalization experts has little impact, whereas the compatibility expert has a greater influence. 
Further investigation revealed that OutfitGAN's scores are relatively extreme, indicating poor generalization performance.
This demonstrates that improving the quality of the expert can further enhance the model's performance.
% The more fashion knowledge the fashion expert possesses, the more accurate their judgments will be, leading to better improvements in the corresponding aspects of the fine-tuned model.

% For more details of the  Alternative Implementations of Experts, please refer to Appendix \ref{sec:expert}.

We further tested the quality of the multi-expert feedback module to ensure that the scores given by these experts are comprehensive and objective.
% As shown in Table \ref{tab:test_multi_expert}, we have set up three evaluation metrics: Accuracy, Mean, and Pair-t. Bold indicates the experts in FashionDPO. Specifically, Accuracy represents the percentage of correctly identified images that match user preferences on the test set. Mean refers to the average score after normalization, and Pair-t indicates the performance difference between the two models on the same test set. 
% To be noted, there is currently a lack of annotated clothing datasets that align with human preferences. Therefore, we use generated fashion item images and the iFashion dataset to construct a test subset which includes positive and negative item pairs. 
As shown in Table 4, in terms of Quality, MiniCPM outperforms GPT-4 in accuracy. This is because GPT-4 has a weaker ability to distinguish between positive and negative fashion items compared to MiniCPM, as evidenced by GPT-4 assigning relatively high scores to negative items as well. 
In terms of Compatibility, the Accuracy of VBPR is significantly higher than that of Outfit-GAN. 
This is because Outfit-GAN's scoring tends to be more extreme, resulting in a lack of distinction. 
In terms of Personalization, both CLIP-Score and VBPR perform similarly in accuracy, and their efficiency in model fine-tuning is also comparable.
From the Pair-t test, it is evident that the p-values between the models for all three evaluation aspects are below the set significance level $(0.05)$, indicating that there are differences between the models. 
It further demonstrates the versatility of the FashionDPO framework, showing that the experts within it are interchangeable, and that stronger experts lead to better fine-tuning performance.
% To be noted, our key contribution is proposing a fine-tuning framework using DPO, in which the experts can provide accurate and comprehensive evaluation results.
% As expert models evolve, this framework can replace them with stronger models or be extended to other evaluation aspects to improve corresponding performance.

% If the result is below the set significance level $(0.05)$, it suggests that there is a significant difference between the two models.

\begin{figure}[t]
    \centering
    \setlength{\abovecaptionskip}{0.1cm} 
    \setlength{\belowcaptionskip}{-0.3cm} 
    % \subfloat{
    % \label{multi_size_is}
    % \includegraphics[width=0.28\linewidth]{images/model_beta_is.pdf}
    % }
    % \hspace{-0.2cm}
    % \subfloat{
    % \label{multi_size_com}
    % \includegraphics[width=0.28\linewidth]{images/model_beta_com.pdf}
    % }
    % \hspace{-0.2cm}
    % \subfloat{
    % \label{multi_size_per}
    \includegraphics[width=0.85\linewidth]{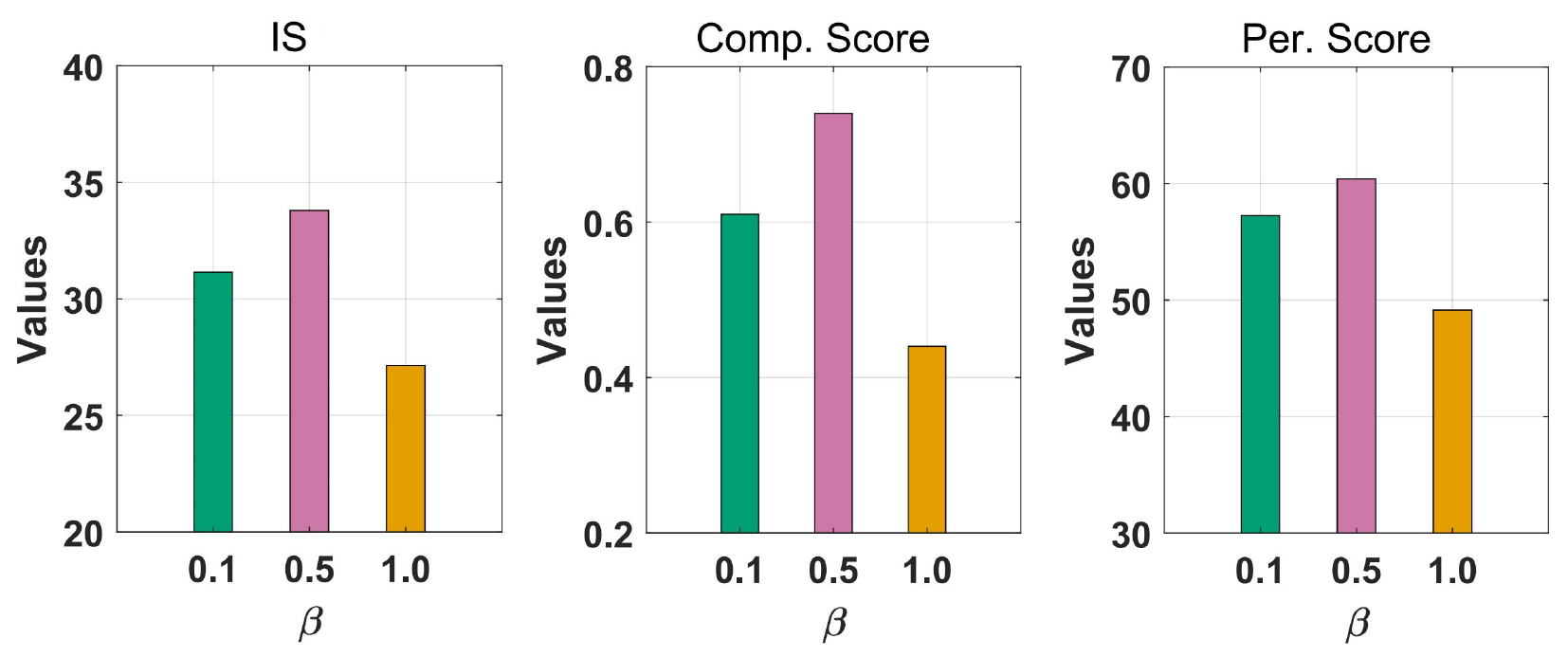}
    \caption{Effects of the hyper-parameter $\beta$ in controlling the deviation. Bars show the scores on the evaluation metric after fine-tuning for five epochs using different hyper-parameters.}
    \label{multi_beta}
\end{figure}

% \begin{figure}[t]
%     \centering
%     \setlength{\abovecaptionskip}{0.1cm} 
%     \setlength{\belowcaptionskip}{-0.4cm} 
%     \subfloat{
%     \label{multi_size_is}
%     \includegraphics[width=0.28\linewidth]{images/model_beta_is.pdf}
%     }
%     \hspace{-0.2cm}
%     \subfloat{
%     \label{multi_size_com}
%     \includegraphics[width=0.28\linewidth]{images/model_beta_com.pdf}
%     }
%     \hspace{-0.2cm}
%     \subfloat{
%     \label{multi_size_per}
%     \includegraphics[width=0.28\linewidth]{images/model_beta_per.pdf}
%     }
%     \caption[]{Effects of the hyper-parameter $\beta$ in controlling the deviation. Bars show the scores on the evaluation metric after fine-tuning for 5 epochs using different hyper-parameters.}
%     \label{multi_beta}
% \end{figure}

% From Figure \ref{multi_experts}, we can see that the two models perform similarly in terms of quality and personalization, as the scoring distributions of the models are similar. For compatibility, we initially used OutfitGAN as the expert; however, training instability occurred during fine-tuning. Further investigation revealed that the scoring distribution of OutfitGAN formed a U-shape within the (0, 1) interval. As a result, we replaced OutfitGAN with VBPR as the compatibility expert in the model. This demonstrates that improving the quality of the expert can further enhance the model's performance, thereby validating the effectiveness of the fine-tuning framework.

\subsubsection{\textbf{Hyper-parameter Analysis}}
The hyper-parameter $\beta = \beta_w = \beta_l$ is used to control the preference and non-preference biases. When $\beta$ increases, it amplifies the differences between preferences and accelerates the model's convergence.  
We fine-tune FashionDPO for five epochs with different $\beta$ and test it on three major evaluation perspectives.
The performance regarding various $\beta$ is shown in Figure \ref{multi_beta}. 
We can see that when $\beta$ increases from $0.1$ to $1.0$, the performance experience a pattern of first growing then dropping.
When $\beta$ is too small, it becomes difficult to distinguish the differences between preferred and non-preferred data, thereby affecting the model's efficiency in learning preferences. When is $\beta$ too large, it overly amplifies the preference differences, which can cause the model to fall into a local optimum.
Therefore, we chose hyper-parameter $\beta = 0.5$.

% \begin{table*}[h]
% \centering
% %\caption{Performance comparison across different methods for PFITB and GOR datasets.}
% \setlength{\tabcolsep}{3pt}
% \begin{tabular}{lcccccccccccccccc}

% 
% \end{tabular}
% \end{table*}

% \begin{figure*}
%   \centering
%   \includegraphics[width=0.95\textwidth]{images/result_multi_size.pdf} 
%   \caption{
%    Results - Multiple dataset sizes
%   }
%   \label{fig:model_study_vbpr}
% \end{figure*}

% \begin{figure}
%   \centering
%   \includegraphics[width=0.45\textwidth]{images/beta.pdf} 
%   \caption{
%     Effects of Beta
%   }
%   \label{fig:result_epochs}
% \end{figure}

\section{Conclusion And Future Work}

In summary, we identified the limitations of lacking diversity and supervised learning paradigm in GOR models, and we propose to leverage AI feedback to address the problem. To address the challenge of the design of AI evaluation models and the mechanisms for updating feedback, we proposed a novel framework FashionDPO, which fine-tunes the fashion outfit generation model using multiple automated experts' feedbacks. 
% We conducted extensive experiments on the iFashion and Polyvore-U datasets for the PFITB and GOR tasks, validating the effectiveness of the fine-tuning framework.

For future work, we will introduce more expert perspectives and explore the interactions among different experts. 
Second, regarding the feedback mechanism, we will incorporate the intensity of preferences to refine the distinctions between preferences and non-preferences. This will enhance our framework's ability to learn expert knowledge.
Furthermore, we will explore additional feedback mechanisms, such as integrating verbal feedback from LLMs into the generation model. 
% Furthermore, we will investigate how multimodal feedback can enhance the model’s overall performance.
% 

% Regarding the feedback mechanism, we will incorporate the intensity of preferences to refine the distinctions between preferences and non-preferences, enhancing the fine-tune framework's ability to learn fashion domain knowledge.

% \input{sec/rebuttal_www}

\newpage
\bibliographystyle{ACM-Reference-Format}
\balance
\bibliography{main}


\begin{thebibliography}{52}

%%% the Web address.

\ifx \showCODEN    \undefined \def \showCODEN     #1{\unskip}     \fi
\ifx \showISBNx    \undefined \def \showISBNx     #1{\unskip}     \fi
\ifx \showISBNxiii \undefined \def \showISBNxiii  #1{\unskip}     \fi
\ifx \showISSN     \undefined \def \showISSN      #1{\unskip}     \fi
\ifx \showLCCN     \undefined \def \showLCCN      #1{\unskip}     \fi
\ifx \shownote     \undefined \def \shownote      #1{#1}          \fi
\ifx \showarticletitle \undefined \def \showarticletitle #1{#1}   \fi
\ifx \showURL      \undefined \def \showURL       {\relax}        \fi
% invisible to TeX
\providecommand\bibfield[2]{#2}
\providecommand\bibinfo[2]{#2}
\providecommand\natexlab[1]{#1}
\providecommand\showeprint[2][]{arXiv:#2}

\bibitem[Chen et~al\mbox{.}(2019)]%
        {POG}
\bibfield{author}{\bibinfo{person}{Wen Chen}, \bibinfo{person}{Pipei Huang}, \bibinfo{person}{Jiaming Xu}, \bibinfo{person}{Xin Guo}, \bibinfo{person}{Cheng Guo}, \bibinfo{person}{Fei Sun}, \bibinfo{person}{Chao Li}, \bibinfo{person}{Andreas Pfadler}, \bibinfo{person}{Huan Zhao}, {and} \bibinfo{person}{Binqiang Zhao}.} \bibinfo{year}{2019}\natexlab{}.
\newblock \showarticletitle{{POG:} Personalized Outfit Generation for Fashion Recommendation at Alibaba iFashion}. In \bibinfo{booktitle}{\emph{{KDD}}}. \bibinfo{publisher}{{ACM}}, \bibinfo{pages}{2662--2670}.
\newblock


\bibitem[Ding et~al\mbox{.}(2024a)]%
        {FashionRecSurvey-23}
\bibfield{author}{\bibinfo{person}{Yujuan Ding}, \bibinfo{person}{Zhihui Lai}, \bibinfo{person}{P.~Y. Mok}, {and} \bibinfo{person}{Tat{-}Seng Chua}.} \bibinfo{year}{2024}\natexlab{a}.
\newblock \showarticletitle{Computational Technologies for Fashion Recommendation: {A} Survey}.
\newblock \bibinfo{journal}{\emph{{ACM} Comput. Surv.}} \bibinfo{volume}{56}, \bibinfo{number}{5} (\bibinfo{year}{2024}), \bibinfo{pages}{121:1--121:45}.
\newblock


\bibitem[Ding et~al\mbox{.}(2024b)]%
        {FashionReGen24}
\bibfield{author}{\bibinfo{person}{Yujuan Ding}, \bibinfo{person}{Yunshan Ma}, \bibinfo{person}{Wenqi Fan}, \bibinfo{person}{Yige Yao}, \bibinfo{person}{Tat{-}Seng Chua}, {and} \bibinfo{person}{Qing Li}.} \bibinfo{year}{2024}\natexlab{b}.
\newblock \showarticletitle{FashionReGen: LLM-Empowered Fashion Report Generation}. In \bibinfo{booktitle}{\emph{{WWW} (Companion Volume)}}. \bibinfo{publisher}{{ACM}}, \bibinfo{pages}{991--994}.
\newblock


\bibitem[Ding et~al\mbox{.}(2022)]%
        {fashion_trend_tmm}
\bibfield{author}{\bibinfo{person}{Yujuan Ding}, \bibinfo{person}{Yunshan Ma}, \bibinfo{person}{Lizi Liao}, \bibinfo{person}{Wai~Keung Wong}, {and} \bibinfo{person}{Tat{-}Seng Chua}.} \bibinfo{year}{2022}\natexlab{}.
\newblock \showarticletitle{Leveraging Multiple Relations for Fashion Trend Forecasting Based on Social Media}.
\newblock \bibinfo{journal}{\emph{{IEEE} Trans. Multim.}}  \bibinfo{volume}{24} (\bibinfo{year}{2022}), \bibinfo{pages}{2287--2299}.
\newblock


\bibitem[Ding et~al\mbox{.}(2023)]%
        {PFOG}
\bibfield{author}{\bibinfo{person}{Yujuan Ding}, \bibinfo{person}{P.~Y. Mok}, \bibinfo{person}{Yunshan Ma}, {and} \bibinfo{person}{Yi Bin}.} \bibinfo{year}{2023}\natexlab{}.
\newblock \showarticletitle{Personalized fashion outfit generation with user coordination preference learning}.
\newblock \bibinfo{journal}{\emph{Inf. Process. Manag.}} \bibinfo{volume}{60}, \bibinfo{number}{5} (\bibinfo{year}{2023}), \bibinfo{pages}{103434}.
\newblock


\bibitem[Dong et~al\mbox{.}(2019)]%
        {personalCom}
\bibfield{author}{\bibinfo{person}{Xue Dong}, \bibinfo{person}{Xuemeng Song}, \bibinfo{person}{Fuli Feng}, \bibinfo{person}{Peiguang Jing}, \bibinfo{person}{Xin{-}Shun Xu}, {and} \bibinfo{person}{Liqiang Nie}.} \bibinfo{year}{2019}\natexlab{}.
\newblock \showarticletitle{Personalized Capsule Wardrobe Creation with Garment and User Modeling}. In \bibinfo{booktitle}{\emph{{ACM} Multimedia}}. \bibinfo{publisher}{{ACM}}, \bibinfo{pages}{302--310}.
\newblock


\bibitem[Du et~al\mbox{.}(2023)]%
        {EBRec}
\bibfield{author}{\bibinfo{person}{Xiaoyu Du}, \bibinfo{person}{Kun Qian}, \bibinfo{person}{Yunshan Ma}, {and} \bibinfo{person}{Xinguang Xiang}.} \bibinfo{year}{2023}\natexlab{}.
\newblock \showarticletitle{Enhancing item-level bundle representation for bundle recommendation}.
\newblock \bibinfo{journal}{\emph{ACM Transactions on Recommender Systems}} (\bibinfo{year}{2023}).
\newblock


\bibitem[Gou et~al\mbox{.}(2023)]%
        {DCI-VTON}
\bibfield{author}{\bibinfo{person}{Junhong Gou}, \bibinfo{person}{Siyu Sun}, \bibinfo{person}{Jianfu Zhang}, \bibinfo{person}{Jianlou Si}, \bibinfo{person}{Chen Qian}, {and} \bibinfo{person}{Liqing Zhang}.} \bibinfo{year}{2023}\natexlab{}.
\newblock \showarticletitle{Taming the Power of Diffusion Models for High-Quality Virtual Try-On with Appearance Flow}. In \bibinfo{booktitle}{\emph{{ACM} Multimedia}}. \bibinfo{publisher}{{ACM}}, \bibinfo{pages}{7599--7607}.
\newblock


\bibitem[Han et~al\mbox{.}(2017)]%
        {withLSTM}
\bibfield{author}{\bibinfo{person}{Xintong Han}, \bibinfo{person}{Zuxuan Wu}, \bibinfo{person}{Yu{-}Gang Jiang}, {and} \bibinfo{person}{Larry~S. Davis}.} \bibinfo{year}{2017}\natexlab{}.
\newblock \showarticletitle{Learning Fashion Compatibility with Bidirectional LSTMs}. In \bibinfo{booktitle}{\emph{{ACM} Multimedia}}. \bibinfo{publisher}{{ACM}}, \bibinfo{pages}{1078--1086}.
\newblock


\bibitem[Han et~al\mbox{.}(2018)]%
        {VITON}
\bibfield{author}{\bibinfo{person}{Xintong Han}, \bibinfo{person}{Zuxuan Wu}, \bibinfo{person}{Zhe Wu}, \bibinfo{person}{Ruichi Yu}, {and} \bibinfo{person}{Larry~S. Davis}.} \bibinfo{year}{2018}\natexlab{}.
\newblock \showarticletitle{{VITON:} An Image-Based Virtual Try-On Network}. In \bibinfo{booktitle}{\emph{{CVPR}}}. \bibinfo{publisher}{Computer Vision Foundation / {IEEE} Computer Society}, \bibinfo{pages}{7543--7552}.
\newblock


\bibitem[He et~al\mbox{.}(2015)]%
        {resnet}
\bibfield{author}{\bibinfo{person}{Kaiming He}, \bibinfo{person}{Xiangyu Zhang}, \bibinfo{person}{Shaoqing Ren}, {and} \bibinfo{person}{Jian Sun}.} \bibinfo{year}{2015}\natexlab{}.
\newblock \showarticletitle{Deep Residual Learning for Image Recognition}.
\newblock \bibinfo{journal}{\emph{CoRR}}  \bibinfo{volume}{abs/1512.03385} (\bibinfo{year}{2015}).
\newblock


\bibitem[He and McAuley(2016)]%
        {VBPR}
\bibfield{author}{\bibinfo{person}{Ruining He} {and} \bibinfo{person}{Julian~J. McAuley}.} \bibinfo{year}{2016}\natexlab{}.
\newblock \showarticletitle{{VBPR:} Visual Bayesian Personalized Ranking from Implicit Feedback}. In \bibinfo{booktitle}{\emph{{AAAI}}}. \bibinfo{publisher}{{AAAI} Press}, \bibinfo{pages}{144--150}.
\newblock


\bibitem[Ho and Salimans(2022)]%
        {CFG}
\bibfield{author}{\bibinfo{person}{Jonathan Ho} {and} \bibinfo{person}{Tim Salimans}.} \bibinfo{year}{2022}\natexlab{}.
\newblock \showarticletitle{Classifier-Free Diffusion Guidance}.
\newblock \bibinfo{journal}{\emph{CoRR}}  \bibinfo{volume}{abs/2207.12598} (\bibinfo{year}{2022}).
\newblock


\bibitem[Huang et~al\mbox{.}(2024)]%
        {PathchDPO}
\bibfield{author}{\bibinfo{person}{Qihan Huang}, \bibinfo{person}{Long Chan}, \bibinfo{person}{Jinlong Liu}, \bibinfo{person}{Wanggui He}, \bibinfo{person}{Hao Jiang}, \bibinfo{person}{Mingli Song}, {and} \bibinfo{person}{Jie Song}.} \bibinfo{year}{2024}\natexlab{}.
\newblock \showarticletitle{PatchDPO: Patch-level {DPO} for Finetuning-free Personalized Image Generation}.
\newblock \bibinfo{journal}{\emph{CoRR}}  \bibinfo{volume}{abs/2412.03177} (\bibinfo{year}{2024}).
\newblock


\bibitem[Huynh et~al\mbox{.}(2018)]%
        {CRAFT}
\bibfield{author}{\bibinfo{person}{Cong~Phuoc Huynh}, \bibinfo{person}{Arri Ciptadi}, \bibinfo{person}{Ambrish Tyagi}, {and} \bibinfo{person}{Amit Agrawal}.} \bibinfo{year}{2018}\natexlab{}.
\newblock \showarticletitle{{CRAFT:} Complementary Recommendations Using Adversarial Feature Transformer}.
\newblock \bibinfo{journal}{\emph{CoRR}}  \bibinfo{volume}{abs/1804.10871} (\bibinfo{year}{2018}).
\newblock


\bibitem[Kim et~al\mbox{.}(2023)]%
        {stableVTON}
\bibfield{author}{\bibinfo{person}{Jeongho Kim}, \bibinfo{person}{Gyojung Gu}, \bibinfo{person}{Minho Park}, \bibinfo{person}{Sunghyun Park}, {and} \bibinfo{person}{Jaegul Choo}.} \bibinfo{year}{2023}\natexlab{}.
\newblock \showarticletitle{StableVITON: Learning Semantic Correspondence with Latent Diffusion Model for Virtual Try-On}.
\newblock \bibinfo{journal}{\emph{CoRR}}  \bibinfo{volume}{abs/2312.01725} (\bibinfo{year}{2023}).
\newblock


\bibitem[Lee and Buu(2024)]%
        {ddpm}
\bibfield{author}{\bibinfo{person}{Hyeon{-}Ju Lee} {and} \bibinfo{person}{Seok{-}Jun Buu}.} \bibinfo{year}{2024}\natexlab{}.
\newblock \showarticletitle{Deep Generative Replay With Denoising Diffusion Probabilistic Models for Continual Learning in Audio Classification}.
\newblock \bibinfo{journal}{\emph{{IEEE} Access}}  \bibinfo{volume}{12} (\bibinfo{year}{2024}), \bibinfo{pages}{134714--134727}.
\newblock


\bibitem[Lee et~al\mbox{.}(2023)]%
        {RLAIF}
\bibfield{author}{\bibinfo{person}{Harrison Lee}, \bibinfo{person}{Samrat Phatale}, \bibinfo{person}{Hassan Mansoor}, \bibinfo{person}{Kellie Lu}, \bibinfo{person}{Thomas Mesnard}, \bibinfo{person}{Colton Bishop}, \bibinfo{person}{Victor Carbune}, {and} \bibinfo{person}{Abhinav Rastogi}.} \bibinfo{year}{2023}\natexlab{}.
\newblock \showarticletitle{{RLAIF:} Scaling Reinforcement Learning from Human Feedback with {AI} Feedback}.
\newblock \bibinfo{journal}{\emph{CoRR}}  \bibinfo{volume}{abs/2309.00267} (\bibinfo{year}{2023}).
\newblock


\bibitem[Lee et~al\mbox{.}(2024)]%
        {RLAIF-sample}
\bibfield{author}{\bibinfo{person}{Harrison Lee}, \bibinfo{person}{Samrat Phatale}, \bibinfo{person}{Hassan Mansoor}, \bibinfo{person}{Thomas Mesnard}, \bibinfo{person}{Johan Ferret}, \bibinfo{person}{Kellie Lu}, \bibinfo{person}{Colton Bishop}, \bibinfo{person}{Ethan Hall}, \bibinfo{person}{Victor Carbune}, \bibinfo{person}{Abhinav Rastogi}, {and} \bibinfo{person}{Sushant Prakash}.} \bibinfo{year}{2024}\natexlab{}.
\newblock \showarticletitle{{RLAIF} vs. {RLHF:} Scaling Reinforcement Learning from Human Feedback with {AI} Feedback}. In \bibinfo{booktitle}{\emph{{ICML}}}. \bibinfo{publisher}{OpenReview.net}.
\newblock


\bibitem[Li et~al\mbox{.}(2020)]%
        {PORGraph}
\bibfield{author}{\bibinfo{person}{Xingchen Li}, \bibinfo{person}{Xiang Wang}, \bibinfo{person}{Xiangnan He}, \bibinfo{person}{Long Chen}, \bibinfo{person}{Jun Xiao}, {and} \bibinfo{person}{Tat{-}Seng Chua}.} \bibinfo{year}{2020}\natexlab{}.
\newblock \showarticletitle{Hierarchical Fashion Graph Network for Personalized Outfit Recommendation}. In \bibinfo{booktitle}{\emph{{SIGIR}}}. \bibinfo{publisher}{{ACM}}, \bibinfo{pages}{159--168}.
\newblock


\bibitem[Liang et~al\mbox{.}(2024)]%
        {SPO}
\bibfield{author}{\bibinfo{person}{Zhanhao Liang}, \bibinfo{person}{Yuhui Yuan}, \bibinfo{person}{Shuyang Gu}, \bibinfo{person}{Bohan Chen}, \bibinfo{person}{Tiankai Hang}, \bibinfo{person}{Ji Li}, {and} \bibinfo{person}{Liang Zheng}.} \bibinfo{year}{2024}\natexlab{}.
\newblock \showarticletitle{Step-aware Preference Optimization: Aligning Preference with Denoising Performance at Each Step}.
\newblock \bibinfo{journal}{\emph{CoRR}}  \bibinfo{volume}{abs/2406.04314} (\bibinfo{year}{2024}).
\newblock


\bibitem[Liu et~al\mbox{.}(2022)]%
        {pndm}
\bibfield{author}{\bibinfo{person}{Luping Liu}, \bibinfo{person}{Yi Ren}, \bibinfo{person}{Zhijie Lin}, {and} \bibinfo{person}{Zhou Zhao}.} \bibinfo{year}{2022}\natexlab{}.
\newblock \showarticletitle{Pseudo Numerical Methods for Diffusion Models on Manifolds}. In \bibinfo{booktitle}{\emph{{ICLR}}}. \bibinfo{publisher}{OpenReview.net}.
\newblock


\bibitem[Liu et~al\mbox{.}(2025)]%
        {BundleMLLM}
\bibfield{author}{\bibinfo{person}{Xiaohao Liu}, \bibinfo{person}{Jie Wu}, \bibinfo{person}{Zhulin Tao}, \bibinfo{person}{Yunshan Ma}, \bibinfo{person}{Yinwei Wei}, {and} \bibinfo{person}{Tat{-}Seng Chua}.} \bibinfo{year}{2025}\natexlab{}.
\newblock \showarticletitle{Fine-tuning Multimodal Large Language Models for Product Bundling}. In \bibinfo{booktitle}{\emph{{KDD} {(1)}}}. \bibinfo{publisher}{{ACM}}, \bibinfo{pages}{848--858}.
\newblock


\bibitem[Lu et~al\mbox{.}(2021)]%
        {PORAnchors}
\bibfield{author}{\bibinfo{person}{Zhi Lu}, \bibinfo{person}{Yang Hu}, \bibinfo{person}{Yan Chen}, {and} \bibinfo{person}{Bing Zeng}.} \bibinfo{year}{2021}\natexlab{}.
\newblock \showarticletitle{Personalized Outfit Recommendation With Learnable Anchors}. In \bibinfo{booktitle}{\emph{{CVPR}}}. \bibinfo{publisher}{Computer Vision Foundation / {IEEE}}, \bibinfo{pages}{12722--12731}.
\newblock


\bibitem[Lu et~al\mbox{.}(2019)]%
        {polyvore_u}
\bibfield{author}{\bibinfo{person}{Zhi Lu}, \bibinfo{person}{Yang Hu}, \bibinfo{person}{Yunchao Jiang}, \bibinfo{person}{Yan Chen}, {and} \bibinfo{person}{Bing Zeng}.} \bibinfo{year}{2019}\natexlab{}.
\newblock \showarticletitle{Learning Binary Code for Personalized Fashion Recommendation}. In \bibinfo{booktitle}{\emph{{CVPR}}}. \bibinfo{publisher}{Computer Vision Foundation / {IEEE}}, \bibinfo{pages}{10562--10570}.
\newblock


\bibitem[Ma et~al\mbox{.}(2020)]%
        {fashion_trend}
\bibfield{author}{\bibinfo{person}{Yunshan Ma}, \bibinfo{person}{Yujuan Ding}, \bibinfo{person}{Xun Yang}, \bibinfo{person}{Lizi Liao}, \bibinfo{person}{Wai~Keung Wong}, {and} \bibinfo{person}{Tat{-}Seng Chua}.} \bibinfo{year}{2020}\natexlab{}.
\newblock \showarticletitle{Knowledge Enhanced Neural Fashion Trend Forecasting}. In \bibinfo{booktitle}{\emph{{ICMR}}}. \bibinfo{publisher}{{ACM}}, \bibinfo{pages}{82--90}.
\newblock


\bibitem[Ma et~al\mbox{.}(2024)]%
        {MultiCBR}
\bibfield{author}{\bibinfo{person}{Yunshan Ma}, \bibinfo{person}{Yingzhi He}, \bibinfo{person}{Xiang Wang}, \bibinfo{person}{Yinwei Wei}, \bibinfo{person}{Xiaoyu Du}, \bibinfo{person}{Yuyangzi Fu}, {and} \bibinfo{person}{Tat{-}Seng Chua}.} \bibinfo{year}{2024}\natexlab{}.
\newblock \showarticletitle{MultiCBR: Multi-view Contrastive Learning for Bundle Recommendation}.
\newblock \bibinfo{journal}{\emph{{ACM} Trans. Inf. Syst.}} \bibinfo{volume}{42}, \bibinfo{number}{4} (\bibinfo{year}{2024}), \bibinfo{pages}{100:1--100:23}.
\newblock


\bibitem[Moosaei et~al\mbox{.}(2022)]%
        {OutfitGAN}
\bibfield{author}{\bibinfo{person}{Maryam Moosaei}, \bibinfo{person}{Yusan Lin}, \bibinfo{person}{Ablaikhan Akhazhanov}, \bibinfo{person}{Huiyuan Chen}, \bibinfo{person}{Fei Wang}, {and} \bibinfo{person}{Hao Yang}.} \bibinfo{year}{2022}\natexlab{}.
\newblock \showarticletitle{OutfitGAN: Learning Compatible Items for Generative Fashion Outfits}. In \bibinfo{booktitle}{\emph{{CVPR} Workshops}}. \bibinfo{publisher}{{IEEE}}, \bibinfo{pages}{2272--2276}.
\newblock


\bibitem[Na et~al\mbox{.}(2024)]%
        {BDPO}
\bibfield{author}{\bibinfo{person}{Sanghyeon Na}, \bibinfo{person}{Yonggyu Kim}, {and} \bibinfo{person}{Hyunjoon Lee}.} \bibinfo{year}{2024}\natexlab{}.
\newblock \showarticletitle{Boost Your Own Human Image Generation Model via Direct Preference Optimization with {AI} Feedback}.
\newblock \bibinfo{journal}{\emph{CoRR}}  \bibinfo{volume}{abs/2405.20216} (\bibinfo{year}{2024}).
\newblock


\bibitem[Parmar et~al\mbox{.}(2022)]%
        {FID}
\bibfield{author}{\bibinfo{person}{Gaurav Parmar}, \bibinfo{person}{Richard Zhang}, {and} \bibinfo{person}{Jun{-}Yan Zhu}.} \bibinfo{year}{2022}\natexlab{}.
\newblock \showarticletitle{On Aliased Resizing and Surprising Subtleties in {GAN} Evaluation}. In \bibinfo{booktitle}{\emph{{CVPR}}}. \bibinfo{publisher}{{IEEE}}, \bibinfo{pages}{11400--11410}.
\newblock


\bibitem[Radford et~al\mbox{.}(2021)]%
        {CLIP}
\bibfield{author}{\bibinfo{person}{Alec Radford}, \bibinfo{person}{Jong~Wook Kim}, \bibinfo{person}{Chris Hallacy}, \bibinfo{person}{Aditya Ramesh}, \bibinfo{person}{Gabriel Goh}, \bibinfo{person}{Sandhini Agarwal}, \bibinfo{person}{Girish Sastry}, \bibinfo{person}{Amanda Askell}, \bibinfo{person}{Pamela Mishkin}, \bibinfo{person}{Jack Clark}, \bibinfo{person}{Gretchen Krueger}, {and} \bibinfo{person}{Ilya Sutskever}.} \bibinfo{year}{2021}\natexlab{}.
\newblock \showarticletitle{Learning Transferable Visual Models From Natural Language Supervision}. In \bibinfo{booktitle}{\emph{{ICML}}} \emph{(\bibinfo{series}{Proceedings of Machine Learning Research}, Vol.~\bibinfo{volume}{139})}. \bibinfo{publisher}{{PMLR}}, \bibinfo{pages}{8748--8763}.
\newblock


\bibitem[Rafailov et~al\mbox{.}(2023)]%
        {DPO}
\bibfield{author}{\bibinfo{person}{Rafael Rafailov}, \bibinfo{person}{Archit Sharma}, \bibinfo{person}{Eric Mitchell}, \bibinfo{person}{Christopher~D. Manning}, \bibinfo{person}{Stefano Ermon}, {and} \bibinfo{person}{Chelsea Finn}.} \bibinfo{year}{2023}\natexlab{}.
\newblock \showarticletitle{Direct Preference Optimization: Your Language Model is Secretly a Reward Model}. In \bibinfo{booktitle}{\emph{NeurIPS}}.
\newblock


\bibitem[Rendle et~al\mbox{.}(2009)]%
        {bpr}
\bibfield{author}{\bibinfo{person}{Steffen Rendle}, \bibinfo{person}{Christoph Freudenthaler}, \bibinfo{person}{Zeno Gantner}, {and} \bibinfo{person}{Lars Schmidt{-}Thieme}.} \bibinfo{year}{2009}\natexlab{}.
\newblock \showarticletitle{{BPR:} Bayesian Personalized Ranking from Implicit Feedback}. In \bibinfo{booktitle}{\emph{{UAI}}}. \bibinfo{publisher}{{AUAI} Press}, \bibinfo{pages}{452--461}.
\newblock


\bibitem[Rombach et~al\mbox{.}(2022)]%
        {SD}
\bibfield{author}{\bibinfo{person}{Robin Rombach}, \bibinfo{person}{Andreas Blattmann}, \bibinfo{person}{Dominik Lorenz}, \bibinfo{person}{Patrick Esser}, {and} \bibinfo{person}{Bj{\"{o}}rn Ommer}.} \bibinfo{year}{2022}\natexlab{}.
\newblock \showarticletitle{High-Resolution Image Synthesis with Latent Diffusion Models}. In \bibinfo{booktitle}{\emph{{CVPR}}}. \bibinfo{publisher}{{IEEE}}, \bibinfo{pages}{10674--10685}.
\newblock


\bibitem[Shih et~al\mbox{.}(2018)]%
        {Compatibility}
\bibfield{author}{\bibinfo{person}{Yong{-}Siang Shih}, \bibinfo{person}{Kai{-}Yueh Chang}, \bibinfo{person}{Hsuan{-}Tien Lin}, {and} \bibinfo{person}{Min Sun}.} \bibinfo{year}{2018}\natexlab{}.
\newblock \showarticletitle{Compatibility Family Learning for Item Recommendation and Generation}. In \bibinfo{booktitle}{\emph{{AAAI}}}. \bibinfo{publisher}{{AAAI} Press}, \bibinfo{pages}{2403--2410}.
\newblock


\bibitem[Silver et~al\mbox{.}(2016)]%
        {RL-1}
\bibfield{author}{\bibinfo{person}{David Silver}, \bibinfo{person}{Aja Huang}, \bibinfo{person}{Chris~J. Maddison}, \bibinfo{person}{Arthur Guez}, \bibinfo{person}{Laurent Sifre}, \bibinfo{person}{George van~den Driessche}, \bibinfo{person}{Julian Schrittwieser}, \bibinfo{person}{Ioannis Antonoglou}, \bibinfo{person}{Vedavyas Panneershelvam}, \bibinfo{person}{Marc Lanctot}, \bibinfo{person}{Sander Dieleman}, \bibinfo{person}{Dominik Grewe}, \bibinfo{person}{John Nham}, \bibinfo{person}{Nal Kalchbrenner}, \bibinfo{person}{Ilya Sutskever}, \bibinfo{person}{Timothy~P. Lillicrap}, \bibinfo{person}{Madeleine Leach}, \bibinfo{person}{Koray Kavukcuoglu}, \bibinfo{person}{Thore Graepel}, {and} \bibinfo{person}{Demis Hassabis}.} \bibinfo{year}{2016}\natexlab{}.
\newblock \showarticletitle{Mastering the game of Go with deep neural networks and tree search}.
\newblock \bibinfo{journal}{\emph{Nat.}} \bibinfo{volume}{529}, \bibinfo{number}{7587} (\bibinfo{year}{2016}), \bibinfo{pages}{484--489}.
\newblock


\bibitem[Silver et~al\mbox{.}(2017)]%
        {RL-2}
\bibfield{author}{\bibinfo{person}{David Silver}, \bibinfo{person}{Julian Schrittwieser}, \bibinfo{person}{Karen Simonyan}, \bibinfo{person}{Ioannis Antonoglou}, \bibinfo{person}{Aja Huang}, \bibinfo{person}{Arthur Guez}, \bibinfo{person}{Thomas Hubert}, \bibinfo{person}{Lucas Baker}, \bibinfo{person}{Matthew Lai}, \bibinfo{person}{Adrian Bolton}, \bibinfo{person}{Yutian Chen}, \bibinfo{person}{Timothy~P. Lillicrap}, \bibinfo{person}{Fan Hui}, \bibinfo{person}{Laurent Sifre}, \bibinfo{person}{George van~den Driessche}, \bibinfo{person}{Thore Graepel}, {and} \bibinfo{person}{Demis Hassabis}.} \bibinfo{year}{2017}\natexlab{}.
\newblock \showarticletitle{Mastering the game of Go without human knowledge}.
\newblock \bibinfo{journal}{\emph{Nat.}} \bibinfo{volume}{550}, \bibinfo{number}{7676} (\bibinfo{year}{2017}), \bibinfo{pages}{354--359}.
\newblock


\bibitem[Song et~al\mbox{.}(2021)]%
        {DDIM}
\bibfield{author}{\bibinfo{person}{Jiaming Song}, \bibinfo{person}{Chenlin Meng}, {and} \bibinfo{person}{Stefano Ermon}.} \bibinfo{year}{2021}\natexlab{}.
\newblock \showarticletitle{Denoising Diffusion Implicit Models}. In \bibinfo{booktitle}{\emph{{ICLR}}}. \bibinfo{publisher}{OpenReview.net}.
\newblock


\bibitem[Szegedy et~al\mbox{.}(2016)]%
        {Inception-V3}
\bibfield{author}{\bibinfo{person}{Christian Szegedy}, \bibinfo{person}{Vincent Vanhoucke}, \bibinfo{person}{Sergey Ioffe}, \bibinfo{person}{Jonathon Shlens}, {and} \bibinfo{person}{Zbigniew Wojna}.} \bibinfo{year}{2016}\natexlab{}.
\newblock \showarticletitle{Rethinking the Inception Architecture for Computer Vision}. In \bibinfo{booktitle}{\emph{{CVPR}}}. \bibinfo{publisher}{{IEEE} Computer Society}, \bibinfo{pages}{2818--2826}.
\newblock


\bibitem[Wallace et~al\mbox{.}(2023)]%
        {Diffusion-DPO}
\bibfield{author}{\bibinfo{person}{Bram Wallace}, \bibinfo{person}{Meihua Dang}, \bibinfo{person}{Rafael Rafailov}, \bibinfo{person}{Linqi Zhou}, \bibinfo{person}{Aaron Lou}, \bibinfo{person}{Senthil Purushwalkam}, \bibinfo{person}{Stefano Ermon}, \bibinfo{person}{Caiming Xiong}, \bibinfo{person}{Shafiq Joty}, {and} \bibinfo{person}{Nikhil Naik}.} \bibinfo{year}{2023}\natexlab{}.
\newblock \showarticletitle{Diffusion Model Alignment Using Direct Preference Optimization}.
\newblock \bibinfo{journal}{\emph{CoRR}}  \bibinfo{volume}{abs/2311.12908} (\bibinfo{year}{2023}).
\newblock


\bibitem[Wei et~al\mbox{.}(2023)]%
        {ELITE}
\bibfield{author}{\bibinfo{person}{Yuxiang Wei}, \bibinfo{person}{Yabo Zhang}, \bibinfo{person}{Zhilong Ji}, \bibinfo{person}{Jinfeng Bai}, \bibinfo{person}{Lei Zhang}, {and} \bibinfo{person}{Wangmeng Zuo}.} \bibinfo{year}{2023}\natexlab{}.
\newblock \showarticletitle{{ELITE:} Encoding Visual Concepts into Textual Embeddings for Customized Text-to-Image Generation}. In \bibinfo{booktitle}{\emph{{ICCV}}}. \bibinfo{publisher}{{IEEE}}, \bibinfo{pages}{15897--15907}.
\newblock


\bibitem[Xie et~al\mbox{.}(2023)]%
        {GP-VTON}
\bibfield{author}{\bibinfo{person}{Zhenyu Xie}, \bibinfo{person}{Zaiyu Huang}, \bibinfo{person}{Xin Dong}, \bibinfo{person}{Fuwei Zhao}, \bibinfo{person}{Haoye Dong}, \bibinfo{person}{Xijin Zhang}, \bibinfo{person}{Feida Zhu}, {and} \bibinfo{person}{Xiaodan Liang}.} \bibinfo{year}{2023}\natexlab{}.
\newblock \showarticletitle{{GP-VTON:} Towards General Purpose Virtual Try-On via Collaborative Local-Flow Global-Parsing Learning}. In \bibinfo{booktitle}{\emph{{CVPR}}}. \bibinfo{publisher}{{IEEE}}, \bibinfo{pages}{23550--23559}.
\newblock


\bibitem[Xu et~al\mbox{.}(2024a)]%
        {DiFashion}
\bibfield{author}{\bibinfo{person}{Yiyan Xu}, \bibinfo{person}{Wenjie Wang}, \bibinfo{person}{Fuli Feng}, \bibinfo{person}{Yunshan Ma}, \bibinfo{person}{Jizhi Zhang}, {and} \bibinfo{person}{Xiangnan He}.} \bibinfo{year}{2024}\natexlab{a}.
\newblock \showarticletitle{Diffusion Models for Generative Outfit Recommendation}. In \bibinfo{booktitle}{\emph{{SIGIR}}}. \bibinfo{publisher}{{ACM}}, \bibinfo{pages}{1350--1359}.
\newblock


\bibitem[Xu et~al\mbox{.}(2024b)]%
        {lora}
\bibfield{author}{\bibinfo{person}{Yuhui Xu}, \bibinfo{person}{Lingxi Xie}, \bibinfo{person}{Xiaotao Gu}, \bibinfo{person}{Xin Chen}, \bibinfo{person}{Heng Chang}, \bibinfo{person}{Hengheng Zhang}, \bibinfo{person}{Zhengsu Chen}, \bibinfo{person}{Xiaopeng Zhang}, {and} \bibinfo{person}{Qi Tian}.} \bibinfo{year}{2024}\natexlab{b}.
\newblock \showarticletitle{QA-LoRA: Quantization-Aware Low-Rank Adaptation of Large Language Models}. In \bibinfo{booktitle}{\emph{{ICLR}}}. \bibinfo{publisher}{OpenReview.net}.
\newblock


\bibitem[Yang et~al\mbox{.}(2023)]%
        {D3PO}
\bibfield{author}{\bibinfo{person}{Kai Yang}, \bibinfo{person}{Jian Tao}, \bibinfo{person}{Jiafei Lyu}, \bibinfo{person}{Chunjiang Ge}, \bibinfo{person}{Jiaxin Chen}, \bibinfo{person}{Qimai Li}, \bibinfo{person}{Weihan Shen}, \bibinfo{person}{Xiaolong Zhu}, {and} \bibinfo{person}{Xiu Li}.} \bibinfo{year}{2023}\natexlab{}.
\newblock \showarticletitle{Using Human Feedback to Fine-tune Diffusion Models without Any Reward Model}.
\newblock \bibinfo{journal}{\emph{CoRR}}  \bibinfo{volume}{abs/2311.13231} (\bibinfo{year}{2023}).
\newblock


\bibitem[Yang et~al\mbox{.}(2018)]%
        {yang2018recommendation}
\bibfield{author}{\bibinfo{person}{Zilin Yang}, \bibinfo{person}{Zhuo Su}, \bibinfo{person}{Yang Yang}, {and} \bibinfo{person}{Ge Lin}.} \bibinfo{year}{2018}\natexlab{}.
\newblock \showarticletitle{From recommendation to generation: A novel fashion clothing advising framework}. In \bibinfo{booktitle}{\emph{2018 7th International Conference on Digital Home (ICDH)}}. IEEE, \bibinfo{pages}{180--186}.
\newblock


\bibitem[Yao et~al\mbox{.}(2024)]%
        {minicpm}
\bibfield{author}{\bibinfo{person}{Yuan Yao}, \bibinfo{person}{Tianyu Yu}, \bibinfo{person}{Ao Zhang}, \bibinfo{person}{Chongyi Wang}, \bibinfo{person}{Junbo Cui}, \bibinfo{person}{Hongji Zhu}, \bibinfo{person}{Tianchi Cai}, \bibinfo{person}{Haoyu Li}, \bibinfo{person}{Weilin Zhao}, \bibinfo{person}{Zhihui He}, {et~al\mbox{.}}} \bibinfo{year}{2024}\natexlab{}.
\newblock \showarticletitle{MiniCPM-V: A GPT-4V Level MLLM on Your Phone}.
\newblock \bibinfo{journal}{\emph{arXiv preprint arXiv:2408.01800}} (\bibinfo{year}{2024}).
\newblock


\bibitem[Yu et~al\mbox{.}(2024)]%
        {HMaVTON}
\bibfield{author}{\bibinfo{person}{Mingzhe Yu}, \bibinfo{person}{Yunshan Ma}, \bibinfo{person}{Lei Wu}, \bibinfo{person}{Kai Cheng}, \bibinfo{person}{Xue Li}, \bibinfo{person}{Lei Meng}, {and} \bibinfo{person}{Tat{-}Seng Chua}.} \bibinfo{year}{2024}\natexlab{}.
\newblock \showarticletitle{Smart Fitting Room: {A} One-stop Framework for Matching-aware Virtual Try-On}. In \bibinfo{booktitle}{\emph{{ICMR}}}. \bibinfo{publisher}{{ACM}}, \bibinfo{pages}{184--192}.
\newblock


\bibitem[Zeng et~al\mbox{.}(2024)]%
        {Jedi}
\bibfield{author}{\bibinfo{person}{Yu Zeng}, \bibinfo{person}{Vishal~M. Patel}, \bibinfo{person}{Haochen Wang}, \bibinfo{person}{Xun Huang}, \bibinfo{person}{Ting{-}Chun Wang}, \bibinfo{person}{Ming{-}Yu Liu}, {and} \bibinfo{person}{Yogesh Balaji}.} \bibinfo{year}{2024}\natexlab{}.
\newblock \showarticletitle{JeDi: Joint-Image Diffusion Models for Finetuning-Free Personalized Text-to-Image Generation}. In \bibinfo{booktitle}{\emph{{CVPR}}}. \bibinfo{publisher}{{IEEE}}, \bibinfo{pages}{6786--6795}.
\newblock


\bibitem[Zhan et~al\mbox{.}(2022)]%
        {A-FKG}
\bibfield{author}{\bibinfo{person}{Huijing Zhan}, \bibinfo{person}{Jie Lin}, \bibinfo{person}{Kenan~Emir Ak}, \bibinfo{person}{Boxin Shi}, \bibinfo{person}{Ling{-}Yu Duan}, {and} \bibinfo{person}{Alex~C. Kot}.} \bibinfo{year}{2022}\natexlab{}.
\newblock \showarticletitle{{\textdollar}A{\^{}}3{\textdollar}-FKG: Attentive Attribute-Aware Fashion Knowledge Graph for Outfit Preference Prediction}.
\newblock \bibinfo{journal}{\emph{{IEEE} Trans. Multim.}}  \bibinfo{volume}{24} (\bibinfo{year}{2022}), \bibinfo{pages}{819--831}.
\newblock


\bibitem[Zhang et~al\mbox{.}(2023)]%
        {controlNet}
\bibfield{author}{\bibinfo{person}{Lvmin Zhang}, \bibinfo{person}{Anyi Rao}, {and} \bibinfo{person}{Maneesh Agrawala}.} \bibinfo{year}{2023}\natexlab{}.
\newblock \showarticletitle{Adding Conditional Control to Text-to-Image Diffusion Models}. In \bibinfo{booktitle}{\emph{{ICCV}}}. \bibinfo{publisher}{{IEEE}}, \bibinfo{pages}{3813--3824}.
\newblock


\bibitem[Zhang et~al\mbox{.}(2018)]%
        {LPIPS}
\bibfield{author}{\bibinfo{person}{Richard Zhang}, \bibinfo{person}{Phillip Isola}, \bibinfo{person}{Alexei~A. Efros}, \bibinfo{person}{Eli Shechtman}, {and} \bibinfo{person}{Oliver Wang}.} \bibinfo{year}{2018}\natexlab{}.
\newblock \showarticletitle{The Unreasonable Effectiveness of Deep Features as a Perceptual Metric}. In \bibinfo{booktitle}{\emph{{CVPR}}}. \bibinfo{publisher}{Computer Vision Foundation / {IEEE} Computer Society}, \bibinfo{pages}{586--595}.
\newblock


\end{thebibliography}

\newpage
\appendix

\end{document}